\documentclass[a4paper,11pt]{article}
\pdfoutput=1 
     \newcommand{\ben}{\begin{equation}} \newcommand{\een}{\end{equation}}
    
\usepackage{jheppub}

\usepackage[utf8x]{inputenc}
\usepackage{color,accents,stmaryrd}
\usepackage{amssymb,upgreek}
\usepackage{amsmath}
\usepackage{indentfirst}

\newcommand{\cl}{C \kern -0.1em \ell}

\newcommand{\be}{\begin{equation}}
 \newcommand{\ee}{\end{equation}}
 \newcommand{\bea}{\begin{eqnarray}}
 \newcommand{\eea}{\end{eqnarray}}

\def\0{\bm0}

 \newcommand{\M}{\mathcal{M}}

\usepackage{graphicx}
\usepackage{bbm}
\usepackage{multirow}
\usepackage{hhline}
\usepackage{float}
\usepackage{tikz}

\long\def\symbolfootnote[#1]#2{\begingroup%
\def\thefootnote{\fnsymbol{footnote}}\footnote[#1]{#2}\endgroup}
\usetikzlibrary{arrows,calc,shapes,decorations.pathreplacing}
\usepackage{caption}
\usepackage{capt-of}
\usepackage{faktor}
\usepackage{pdfpages}
\usepackage{url}
\usepackage[all]{xy}

\usepackage{hyperref}
\usepackage{xcolor}

\newcommand{\beq}{\begin{eqnarray}}
\newcommand{\eeq}{\end{eqnarray}}

\newcommand{\blkdiam}{\tikz\node[rectangle, draw, yscale=1, scale=.47, rotate=45, fill=black] {};}

\newcommand{\cliff}{{\mathcal{C}\ell}_{p, q}}

\newcommand{\tangb}{T^{*}\mathcal{M}}

\newcommand{\extbund}{{{\textstyle \bigwedge}}(T^{*}\mathcal{M})}

\newcommand{\extfib}{{{\textstyle \bigwedge}}(T^{*}_x \mathcal{M})}		

\newcommand{\extevfib}{{{\textstyle \bigwedge}}^+(T^{*}_x \mathcal{M})}

\newcommand{\secextabrev}{\varGamma({{\textstyle \bigwedge}})}

\newcommand{\secextkabrev}{\varGamma({{\textstyle \bigwedge}}^k)}

\newcommand{\secext}{\varGamma({{\textstyle \bigwedge}} (T^{*}\mathcal{M}))}

\newcommand{\unsecex}{1_{\varGamma}}

\newcommand{\secp}{\varGamma(\mathcal{P})}

\newcommand{\secpe}{\varGamma(\mathcal{P}^+)}

\newcommand{\secpm}{\varGamma(\mathcal{P}^\pm)}

\definecolor{clearblue}{rgb}{0,0.5,0.9}
\definecolor{orange}{rgb}{1,0.5,0}

\title{\boldmath New spinor classes on the Graf--Clifford algebra}

\author[a]{R. Lopes}
\author[b]{R. da Rocha}

\affiliation[a,b]{Centro de Matem\'atica, Computa\c c\~ao e Cogni\c c\~ao, Universidade Federal do ABC - UFABC,  09210-580, Santo Andr\'e, Brazil.}
\emailAdd{rian.lopes@ufabc.edu.br}
\emailAdd{roldao.rocha@ufabc.edu.br}
\abstract{Pinor and spinor fields are  
	sections of the subbundles whose fibers are the representation spaces of the Clifford algebra of the forms, equipped with the Graf product. In this context, pinors and spinors are here considered and the geometric generalized Fierz identities provide the necessary framework to derive and construct new spinor classes on the space of smooth sections of the exterior bundle, endowed with the Graf product, for prominent specific signatures, whose applications are discussed. }
\begin{document}
	\maketitle
	\flushbottom

	\section{Introduction}
	The Clifford algebras classification provides a relationship between supersymmetry and division algebras \cite{oxford,Bonora:2009ta}. 
	From the classical point of view, spinors can be defined as objects which carry an irreducible representation of the Spin group, which is the double covering of the special orthogonal group. Therefore spinors carry the spin 1/2 representation of the group of rotations in a quadratic space. On the other hand, the Spin group  is naturally embedded into a Clifford algebra and an ulterior, equivalent, definition of spinor can be introduced, namely, the algebraic one. In fact, the representation space associated with an irreducible regular representation is a minimal
	left ideal related to the Clifford algebra \cite{lou2}. An algebraic spinor is an element of a minimal left ideal in a Clifford algebra. The representation of the Clifford algebra obtained is called a spinor representation. 
	Classical spinors can be classified with respect to their  bilinear covariants, satisfying the generalized Fierz identities presented in Refs. \cite{bab1, bilfierz}. This property has led to the well known Lounesto's classification of spinors in Minkowski spacetime $\mathbb{R}^{1,3}$ into six classes \cite{lou2}, into regular and singular spinors \cite{Fabbri:2014zya,fabbri,Fabbri:2014wda,Vignolo:2011qt,hoff}. {\color{black}{Prominent features of singular spinors were studied in Ref. \cite{Rogerio:2017gvr}}}. The spinors themselves can be reconstructed from their spinor bilinear covariants, by the reconstruction theorem  \cite{Cra}, {\color{black}{yielding a reciprocal spinor field classification \cite{Cavalcanti:2014wia,fabbri}}}.
	
	{\color{black}{The geometric Fierz identities are constraints on the bilinear covariants, yielding to classify spinors fields on any simply connected manifold with spin structure   \cite{bab1}. Recently, a classification of spinor fields on  Lorentzian \cite{Bon14} and Riemannian   \cite{Bonora:2015ppa} 7-manifolds has been constructed, emulating the Lounesto's classification. It introduces new classes of spinors and new fermionic solutions in supergravity, regarding, in particular, its AdS$_5\times S^5$ \cite{brito,Mendes:2017hmv} and AdS$_4\times S^7$  \cite{Bon14} compactifications. }}
	
	Having constructed new classes of spinors, implicitly using the Graf product and the geometric Fierz identities \cite{bab1}, for signatures $(7, 0)$, $(6,1)$, and $(1, 4)$  \cite{Bon14,Bonora:2015ppa,brito}, our main aim here is to derive  new classes of spinors on a Lorentzian space with signature $(1, 2)$, employing this method, and also to construct pinors in signature $(9,0)$. This last case is naturally linked to spinors in $\mathbb{R}^{1,9}$, as elements that carry the irreducible representation of the group Spin$_{1,9}$, whose Lie algebra $\mathfrak{so}(1,9) \simeq \mathfrak{spin}(1,9)$ is isomorphic to the Lorentz-like Lie algebra $\mathfrak{sl}(2,\mathbb{O})$. Since the Fierz identities can be used to provide 
	quite few realisations, we shall discuss here the last mod 8 possibility \cite{bab1}, by studying the occurrence of new 
	pinor and spinor fields on manifolds of signatures $(1,2)$ and $(9,0)$. This last case is relevant in either the cone or the cylinder formalism, to study  compactifications of $M$-theory with one supersymmetry, lifting 
	the formalism on an 8-dimensional manifold onto a 9-dimensional conic or cylindric metric space, wherein the pin/spin bundle is a real vector bundle \cite{bab2}. Hence, it is important to derive new classes of pinors in these signatures. 
	
	This paper is organized as follows: after presenting the fundamental properties of Clifford bundles in Sect. II, the Graf product is then introduced and studied in Sect. III, in the context of the Graf--Clifford algebra. 
	In Sect. IV, pinors, spinors and the geometric Fierz identities are then constructed, from the Atiyah--Bott--Shapiro mod 8 periodicity. The main subalgebra is a pivotal ingredient for analysing the normal, the almost complex, and the quaternionic  cases for the geometric Fierz identities, in the context of the Clifford bundles. In Sect. V, three non-trivial new classes of spinors on a 3-dimensional manifold of signature $(1,2)$ are constructed, in a detailed and systematic analysis of the geometric Fierz identities. In Sect. VI, seven non-trivial pinor classes of Euclidean 9-manifolds are derived. Sect. VII is finally devoted for 
	drawing the conclusions and perspectives. 
	
	\section{Preliminaries}
	
	Let us regard the 2-uple $(\mathcal{M}, g)$ as a pseudo-Riemannian manifold with signature $(p, q)$, the exterior bundle $\extbund$ and $U$ an open subset on $\mathcal{M}$. The $k$-forms are defined as local sections of the $k$-power exterior bundle  ${\textstyle \bigwedge}^{k}(T^{*}\mathcal{M})$, for $k = 0, \ldots, \dim \mathcal{M} $. The function $1_{\varGamma} \in C^{\infty}(\mathcal{M})$ denotes the unit element of $\varGamma(U, {\textstyle \bigwedge}(T^{*} \mathcal{M}))$. The following suitable notations $\varGamma({{\textstyle \bigwedge}}^k (T^{*}\mathcal{M})) = \varGamma({{\textstyle \bigwedge}}^k)$ and $\varGamma({{\textstyle \bigwedge}} (T^{*}\mathcal{M})) = \varGamma({{\textstyle \bigwedge}})$ shall be adopted, denoting the space of smooth sections of the exterior bundle.  
	If $ \dim\mathcal{M} = n$, given a set of indexes $I_n=\{i_1, \ldots, i_n\}$, let $\{ e_i\;|\; i \in I_n \}$ be a local frame to the tangent bundle $\tangb$ on  $U \subset \mathcal{M}$. A coframe for $T^{*} \mathcal{M}$ is constituted by the set of covectorial sections $\{e^i\;|\; i \in I\}$ that are defined with respect to the dual relation $e^i (e_j) = \delta^{i}_{j} \unsecex$. The metric tensor $g^{*}:\varGamma(U, T^{*} \mathcal{M}) \times \varGamma(U, T^{*} \mathcal{M})\to\mathbb{R}$ is such that $g^{*}(e^i, e^j) =: g^{ij}$.
	The $k$-forms $e^{I_j} = e^{i_1 \ldots i_j} = e^{i_1} \wedge  e^{i_2} \wedge \cdots \wedge  e^{i_j}$ are elements of $\varGamma(U, {\textstyle \bigwedge} (T^{*} \mathcal{M}))$ for $j = 1, \ldots, n$. Therefore, a form $f \in \varGamma(U, {\textstyle \bigwedge} (T^{*} \mathcal{M}))$ reads $
	f = \sum^{n}_{k=1} f_{I_k} e^{I_k},$ 
	where $f_{I_k}$ is a smooth function that constitutes a linear combination of element of $I_k$. Since the fibers of the exterior bundle are $\mathbb{Z}_2$-graded, then there is a splitting ${\textstyle \bigwedge} (T^{*} \mathcal{M}) = {\textstyle \bigwedge}^{+} (T^{*} \mathcal{M}) \oplus {\textstyle \bigwedge}^{-} (T^{*} \mathcal{M})$, where ${\textstyle \bigwedge}^{+} (T^{*} \mathcal{M})$ is the even subbundle whose fibers are given by the grade involution $\#$ as being $\extevfib = \{f \in \extfib \mid \#(f)=f\}$. Besides, there is another $\mathbb{Z}_2$-grading on the sections of exterior bundle, given by the odd subspace  $\varGamma^- ({\textstyle \bigwedge}) := \varGamma ({\textstyle \bigwedge}^-) = \bigoplus_{k=\text{odd}} \secextkabrev = \ker (\# + Id_{\secextabrev})$ and by the even subbundle  $\varGamma^+ ({\textstyle \bigwedge}) := \varGamma ({\textstyle \bigwedge}^+) = \bigoplus_{k=\text{even}} \secextkabrev = \ker (\# - Id_{\secextabrev})$. 
	The volume form $
	\textbf{v} = e^{12\ldots n}$ is defined as an element of the space $\varGamma\left( {{\textstyle \bigwedge}}^{n} (T^{*} \mathcal{M})\right)$. 
	\section{The Graf product}
	
	The Graf product $\diamond$ between a $m$-form $f \in  \varGamma( {{\textstyle \bigwedge}}^m (T^{*} \mathcal{M}))$ and a $r$-form $g \in  \varGamma( {{\textstyle \bigwedge}}^r (T^{*} \mathcal{M}))$, $m \leq r$, \cite{frame,cep,Graf,Linhares:1985xa} is defined as 
	\begin{equation}
		f \diamond g = \displaystyle{\sum^{m}_{k=0}} \frac{1}{k!} (-1)^{k(m-k)+[\frac{k}{2}]} f \wedge_k g, 
	\end{equation}
	and the non-commutative relation 
	\begin{equation}
		g \diamond f = (-1)^{mr} \displaystyle{\sum^{m}_{k=0}} \frac{1}{k!}(-1)^{k(m-k+1)+[\frac{k}{2}]} f \wedge_k g,
	\end{equation}
	where $\wedge_k$ denotes the contracted wedge product between $f$ and $ g$ \cite{cep, Car}, iteratively constructed as:
	\begin{eqnarray}
		f \wedge_0 g &=& f \wedge g,\\
		f \wedge_k g &=& \sum^{n}_{i_k, j_k = 1} g^{i_k j_k} (e_{i_k} \rfloor f) \wedge_{k-1} (e_{j_k} \rfloor g).
	\end{eqnarray}

	In Ref. \cite{frame} it was proved that the set of sections $\secext$ is a Clifford algebra with respect to the Graf product, such algebra $(\secext, \diamond)$ is named the Graf--Clifford algebra.
	Due to this definition,  the  volume element $\textbf{v}$ satisfies \cite{frame}
	\begin{equation}\label{volvol}
		\textbf{v} \diamond \textbf{v} =  \left\{ \begin{array}{ll}
			+ \unsecex, & \ \text{if} \  p-q \equiv_8 0, 1, 4, 5 \\
			- \unsecex, & \ \text{if} \  p-q \equiv_8 2, 3, 6, 7 
		\end{array} \right. ,
	\end{equation}
	{\color{black}{remembering that $p-q$ regards the signature of the pseudo-Riemannian metric that endows the $\mathcal{M}$ manifold.}}
	
	For an arbitrary $r$-form $f$ it implies that 
	\begin{equation}\label{fvol}
		f \diamond \textbf{v} = \frac{1}{r!} (-1)^{[\frac{r}{2}]} f \wedge_r \textbf{v}. 
	\end{equation}
	In addition, the volume element $\textbf{v}$ is central in $ \varGamma({{\textstyle \bigwedge}}(T^{*}\mathcal{M}))$, with the  condition that the dimension of $\mathcal{M}$ is odd whichever the grade of the form \cite{frame}.
	
	Throughout this text the adopted definition  for the Hodge operator $\star$ is
	\begin{equation}
		\begin{array}{cccc} \star : & \varGamma({{\textstyle \bigwedge}}^m(T^{*}\mathcal{M}))  & \rightarrow & \varGamma({{\textstyle \bigwedge}}^{n-m}(T^{*}\mathcal{M})) \\
			& f & \mapsto &  f \diamond \textbf{v}
		\end{array}.
	\end{equation}
	The right regular representation on an element $f \in \varGamma({\textstyle \bigwedge})$ by the element $p_\pm:= \frac{1}{2} (\unsecex \pm \textbf{v})$ is defined by $R_{p_\pm} (f) = P_\pm (f)$, where $P_\pm (f) := f \diamond p_\pm$, reading 
	\begin{equation}
		R_{p_\pm} (f) = \frac{1}{2} (f \pm \star f),
	\end{equation} namely, $P_\pm = \frac{1}{2} (Id_{\varGamma({\textstyle \bigwedge})} \pm \star).$
	
	Let us consider the sets $\varGamma_\pm := P_\pm (\varGamma({\textstyle \bigwedge})) = \varGamma({\textstyle \bigwedge}) \diamond p_\pm$, which are not necessarily subalgebras of $(\secextabrev, \diamond)$. When $p-q \equiv_8 0, 1, 4, 5$, if $f_\pm \in \varGamma_\pm$,  then $\star f_\pm= \pm f_\pm$, yielding the identification 
	$\varGamma_\pm = \{f_\pm \in \secextabrev |  \star f_\pm= \pm f_\pm \}$. Besides, this means that if $f \in \varGamma_\pm$,  thus $
	P_\pm (f) =f.$ Given two arbitrary forms $f, g \in \varGamma({\textstyle \bigwedge})$, it follows that  
	\begin{equation}\label{end}
		\!\!\!\!\!\!\!\!\!\!\!\!\!\!P_\pm (f) \diamond P_\pm (g)=\frac{1}{4} f\!\diamond\!g \pm \frac{1}{4} f\!\diamond\!\textbf{v} \!\diamond\!g  \pm  \frac{1}{4} f\!\diamond\!g\!\diamond\!\textbf{v} + \frac{1}{4} f\!\diamond\!\textbf{v}\!\diamond\!g\!\diamond\!\textbf{v}.
	\end{equation}
	In Eq. \eqref{end}, if $p-q \equiv_8 0, 1, 4, 5$ and  $n$ is odd, then we have  $P_\pm (f \diamond g)=P_\pm (f) \diamond P_\pm (g)$. It  means that the $P_\pm$ represent  endomorphisms of the bundle $\varGamma({\textstyle \bigwedge})$. Therefore, the sets $(\varGamma_\pm, \diamond)$ are subalgebras whose units are given by $P_\pm (\unsecex) =  p_\pm$.
	
	Defining $\varGamma_L = \bigoplus_{j=0}^{[\frac{n}{2}]} \varGamma({{\textstyle \bigwedge}}^j)$ and $\varGamma_U = \bigoplus_{j=[\frac{n}{2}]+1}^{n} \varGamma({{\textstyle \bigwedge}}^j)$, we can consider the splitting
	\begin{equation}
		\varGamma\left({\textstyle \bigwedge}\right) = \varGamma_L\oplus \varGamma_U, 
	\end{equation}
	defining the  upper truncation, $P_U (f)= f_U :=\sum_{j=[\frac{n}{2}]+1}^{n} f_{I_j}^{j} e^{I_j}$, and the lower truncation, $P_L (f)= f_L := \sum_{j=0}^{[\frac{n}{2}]} f_{I_j}^{j} e^{I_j}$. In other words, $\varGamma_L = P_L (\varGamma({\textstyle \bigwedge}))$ and $\varGamma_U = P_U (\varGamma({\textstyle \bigwedge}))$.
	
	For a form $f$ in the signatures $p-q \equiv_8 0, 1, 4, 5$, the Hodge operator action is the identity, up to a sign, $\star f = \pm f$. It implies that $\pm(f_L + f_U )= \star f_L + \star f_U$. Now, as the  $f_L, \star f_U$ are in the set $\bigoplus_{j=0}^{[\frac{n}{2}]} \varGamma({{\textstyle \bigwedge}}^j)$ and, on the other hand, $f_U, \star f_L$ are elements of $ \bigoplus_{j=[\frac{n}{2}]+1}^{n} \varGamma({{\textstyle \bigwedge}}^j)$, then it is possible to assert  that $f_U = \pm \star f_L$ and $f_L = \pm \star{f_U}$. 
	Hence, as $f = f_L + f_U$, then 
	\begin{equation}\label{f2}
		f = f_L \pm \star f_L =   P_\pm (2 P_L(f)).
	\end{equation}
	The image of $\varGamma_L$ is not a subalgebra of $\varGamma({\textstyle \bigwedge})$ when it is equipped with the Graf product. Hence, it motivates the definition of the truncated Graf product $\blkdiam_\pm : \secextabrev \rightarrow \varGamma_L$:
	\begin{equation}\label{P-}
		f \blkdiam_\pm g = 2 P_L (P_\pm (f) \diamond P_\pm (g)), \ \forall f, g \in \secextabrev,
	\end{equation}
	which implies that $P_\pm (f \blkdiam_\pm g) =  P_\pm (f) \diamond P_\pm (g)$. 
	Hence, when $n$ is odd and $p-q \equiv_8 0, 1, 4, 5$, the Eq. \eqref{P-} implies that $f \blkdiam_\pm g = 2 P_L (P_\pm (f \diamond g))$ for all $f, g \in \varGamma_L$, and consequently $P_\pm (f \blkdiam_\pm g) = P_\pm (f \diamond g)$, i. e., $P_\pm$ is a endomorphism.
	
	The isomorphism between the subalgebras $(\varGamma_L, \blkdiam_\pm)$ and $(\varGamma_\pm, \diamond)$ was established in Ref. \cite{frame}. Thus, when $n$ is odd and $p-q \equiv_8 0, 1, 4, 5$, the truncated subalgebra $(\varGamma_L, \blkdiam_\pm)$ will be modeled by the subalgebra $(\varGamma_\pm, \diamond)$.
	
	\vspace*{-0.3cm}
	\section{Pinors, spinors and Fierz identities}
	
	{\color{black}{A way to approach pinors and spinors consists of respectively considering elements that carry the irreducible representation in $\cliff$ or $\cliff^+$, respectively \cite{oxford}}}. In fact, unless $n$ equals 3 or 7 modulo 8, the algebra ${\mathcal{C}\ell}_n$ is a real $\mathbb{R}$, a 
	complex ($\mathbb{C}$), or a quaternionic ($\mathbb{H}$) matrix algebra. Hence,  it has a unique irreducible representation, known as the space of pinors. Since the group Pin($n$) is embedded into the algebra ${\mathcal{C}\ell}_n$, the irreducible representations of ${\mathcal{C}\ell}_n$ then restrict to representations
	of Pin($n$). Similarly, Spin($n$) is embedded into the even subalgebra
	${\mathcal{C}\ell}^+_n$ of ${\mathcal{C}\ell}_n$. Thus the
	irreducible representations of ${\mathcal{C}\ell}^+_n$ restrict to irreducible representations of Spin($n$), called spinors \cite{oxford}. The Clifford algebra representations, coming from the Atiyah--Bott--Shapiro theorem, is comprised below \cite{oxford}. 
	\begin{center}
		{\footnotesize{\begin{tabular}{||c||c|c|c|c||} \hline \hline 
					$\begin{matrix} p-q \\ \text{mod} \, 8 \end{matrix}$ & 0 & 1 & 2 
					& 3 \\ \hline 
					$ {\mathcal{C}\ell}_{p,q} $ & $\mathcal{M}(2^{[n/2]},\mathbb{R})$ & 
					$\begin{matrix} \mathcal{M}(2^{[n/2]},\mathbb{R}) \\ \oplus \\
					\mathcal{M}(2^{[n/2]},\mathbb{R}) \end{matrix}$ & 
					$\mathcal{M}(2^{[n/2]},\mathbb{R}) $ & 
					$\mathcal{M}(2^{[n/2]},\mathbb{C}) $ \\ \hline 
					$\begin{matrix}p-q \\ \text{mod} \, 8\end{matrix}$ & 4 
					& 5 & 6 & 7 \\ \hline 
					${\mathcal{C}\ell}_{p,q}$ & $\mathcal{M}(2^{[n/2]-1},\mathbb{H})$ & 
					$\begin{matrix} \mathcal{M}(2^{[n/2]-1},\mathbb{H}) \\ \oplus \\
					\mathcal{M}(2^{[n/2]-1},\mathbb{H}) \end{matrix} $ & 
					$\mathcal{M}(2^{[n/2]-1},\mathbb{H})$ & 
					$\mathcal{M}(2^{[n/2]},\mathbb{C})$ \\ \hline\hline 	
		\end{tabular}}}\\
		\medskip
		{\footnotesize{{Classification of the real Clifford algebras}, for $p + q = n$.}}
	\end{center}

\textcolor{black}{The Clifford bundle over the cotangent bundle $\mathcal{C}\ell (T^{*}\mathcal{M}) \rightarrow \M$ is defined by the disjoint union as follows
\begin{equation}
\mathcal{C}\ell (T^{*}\mathcal{M}) = \displaystyle{\bigsqcup_{x \in \mathcal{M}}} \mathcal{C}\ell (T^{*}_x \mathcal{M}, g_x),
\end{equation}	
whereas, the even Clifford subbundle ${\mathcal{C}\ell}^{+} (T^{*}\mathcal{M}) \rightarrow \M$, whose fibers are even Clifford algebras, is defined as follows
\begin{equation}
{\mathcal{C}\ell}^{+} (T^{*}\mathcal{M}) = \displaystyle{\bigsqcup_{x \in \mathcal{M}}} {\mathcal{C}\ell}^{+} (T^{*}_x \mathcal{M}, g_x).
\end{equation}}
	
\textcolor{black}{The \textit{pin bundle} $\pi_p: \mathcal{P} \rightarrow \M$ is the bundle whose fibers are the irreducible representation spaces of the fibers $\mathcal{C}\ell (T^{*}_x \mathcal{M}, g_x)$ in $\mathcal{C}\ell (T^{*}\mathcal{M})$, for all $x \in U \subset \M$, this is, considering the irreducible representations $\lambda_x: \mathcal{C}\ell (T^{*}_x \mathcal{M}, g_x) \rightarrow \operatorname{End} P_x$, the pin bundle is given by:
\begin{equation}
\mathcal{P} = \displaystyle{\bigsqcup_{x \in \mathcal{M}}} P_x.
\end{equation}
Thus, the algebra bundle morphism can be considered
\begin{equation}
\lambda : \mathcal{C}\ell (T^{*}\mathcal{M}) \rightarrow \operatorname{End} \mathcal{P},
\end{equation}	
where $\operatorname{End} \mathcal{P}$ is the bundle whose fibers are $\operatorname{End} P_x$ for all $x \in U \subset \M$. A section $\alpha: \M \rightarrow \mathcal{P}$ is named a \textit{pinor field}.}
	
On the other hand, the\textit{ spin bundle} $\pi_s: \mathcal{S} \rightarrow \M$ is the bundle constituted by the irreducible representation spaces of ${\mathcal{C}\ell}^{+} (T^{*}_x \mathcal{M}, g_x)$ in ${\mathcal{C}\ell}^{+} (T^{*}\mathcal{M})$, $x \in U$. This means that the bundle $\mathcal{S}$ is given as follows  
\begin{equation}
\mathcal{S} = \displaystyle{\bigsqcup_{x \in \mathcal{M}}} S_x,
\end{equation}
where each $S_x$ comes from the mapping $\lambda^{+}_x: {\mathcal{C}\ell}^{+} (T^{*}_x \mathcal{M}, g_x) \rightarrow \operatorname{End} S_x$. Naturally, a \textit{spinor field} is defined as being a section of $\mathcal{S}$. Analogously to the pinor bundle case, the following algebra bundle morphism can be defined:
\begin{equation}
\lambda^{+} : {\mathcal{C}\ell}^{+} (T^{*}\mathcal{M}) \rightarrow \operatorname{End} \mathcal{S}.
\end{equation}	
The mappings   $\lambda_x$ and $\lambda^+_x$ are the pinor and spinor representations already defined. For $f \in {\mathcal{C}\ell} (T^{*}_x \mathcal{M}, g_x)$, the mapping $f_x:=\lambda_x(f): P_x \rightarrow P_x$ is a module homomorphism. 

Once the pin and spin bundles are established, consider the following mappings regarding the Graf--Clifford algebra:
	\begin{eqnarray}
		\lambda^\varGamma : \left(\varGamma\left({\textstyle \bigwedge}\right), \diamond\right) &\rightarrow& (\operatorname{End} \varGamma(\mathcal{P}), \circ),\\\label{488}
		\lambda^{\varGamma+}: \left(\varGamma^+\left({{\textstyle \bigwedge}}\right), \diamond\right) &\rightarrow& (\operatorname{End} \varGamma(\mathcal{S}), \circ),
	\end{eqnarray}
	{\color{black}{where ``$\circ$'' denotes the natural  product of endomorphisms.}} Besides, the morphism $\lambda^\varGamma$ satisfies $ \lambda^\varGamma(\unsecex) = Id_{\varGamma(\mathcal{P})}$ and $\lambda^\varGamma(f_1 \diamond f_2) = \lambda^\varGamma (f_1) \circ \lambda^\varGamma(f_2),$ for all $f_1, f_2 \in \varGamma\left( {\textstyle \bigwedge}\right)$.
	
	Observe that the mapping $\lambda^{\varGamma+}$ is the application $\lambda^\varGamma$ restricted to $\varGamma^+\left({{\textstyle \bigwedge}}\right)$, then there exist an identification between $\varGamma(\mathcal{P}^+)$ (when the bundle $\mathcal{P}$ is decomposable) and $\varGamma(\mathcal{S})$. In literature \cite{bab2}, the sections of $\mathcal{P}^\pm$ are named \textit{Majorana-Weyl spinors} when $p-q \equiv_8 0$, sections of $\mathcal{P}^\pm$ are called \textit{symplectic Majorana-Weyl spinors} when $p-q \equiv_8 4$, the sections of $\mathcal{P}^+$ are named \textit{symplectic Majorana spinors} when $p-q \equiv_8 6$ and sections of $\mathcal{P}^+$ are called \textit{Majorana spinors} when $p-q \equiv_8 7$.
		
	If $f \in \varGamma\left( {\textstyle \bigwedge}\right)$, then $f = \displaystyle{\sum^{n}_{k=1}} f_{I_k}^{k} e^{I_k}$, yielding 
	\begin{equation}
		\lambda^\varGamma(f) = \displaystyle{\sum^{n}_{k=1}} f_{I_k}^{k} \lambda^\varGamma(e^{I_k}),
	\end{equation}
	where $ \lambda^\varGamma(e^{I_k}) = \lambda^\varGamma(e^{i_1 \ldots i_k}) = \lambda^\varGamma(e^{i_1} \wedge \cdots \wedge e^{i_k}) = \lambda^\varGamma(e^{i_1}) \circ \cdots \circ \lambda^\varGamma(e^{i_k})$.
	
	Note that the image $\operatorname{Im}\lambda_x$ is whithin the algebra $\operatorname{End} P_x$. By a direct consequence of the Schur's lemma, the algebra $\operatorname{End} P_x$ is a division algebra, since $\lambda_x$ is irreducible.
	
	The real algebra $(\operatorname{End} P_x, \circ)$ is associative. Thus, by the Frobenius theorem for division algebras, each $\operatorname{End} P_x$ is isomorphic either to $\mathbb{R}$, $\mathbb{C}$ or $\mathbb{H}$. Thereat, define the subset
	\begin{equation}
		\mathcal{A}_x:= \operatorname{Cen}_{\operatorname{End} P_x}(\operatorname{Im}\lambda_x).
	\end{equation}
	Since $\operatorname{End} P_x$ is associative, thus $\mathcal{A}_x$ is a subalgebra. 
	Since the centralizer is a division algebra, then $\mathcal{A}_x$ is a division algebra, as $\operatorname{End} P_x$. Namely, $\mathcal{A}_x$ is isomorphic to $\mathbb{R}$, $\mathbb{C}$ or $\mathbb{H}$.
	This subalgebra $\mathcal{A}_x$ is very important to analyze the geometric Fierz identities into three different cases, according to the so called isomorphism type of $\mathcal{A}_x$. Therefore,  this subalgebra will be called hereon the \textit{main subalgebra}.
	With the results by Okubo \cite{oku1}, it is possible to conclude that each element $f_x \in \mathcal{A}_x$ can be written as:\medbreak
	{1) The {normal} case}:
	\begin{equation}
		f_x = a_0 I_{\mathsf{k}_\mathbb{R}}, \ a_0 \in \mathbb{R},
	\end{equation}
	where $I_{\mathsf{k}_\mathbb{R}}$ is the identity matrix of order $\mathsf{k}_\mathbb{R} \times \mathsf{k}_\mathbb{R}$. This means that $\mathcal{A}_x \cong \mathbb{R}$.
	This case occurs when $p-q \equiv_8 0, 1, 2$ and $\mathsf{k}_\mathbb{R} = 2^{[\frac{n}{2}]}$.
	\medbreak
	{2) The {almost complex} case}:
	\begin{equation}
		f_x = a_1 I_{\mathsf{k}_\mathbb{C}} + b_1 J, \ a_1, b_1 \in \mathbb{R}, 
	\end{equation}
	where $J$ is a matrix of order $\mathsf{k}_\mathbb{C} \times \mathsf{k}_\mathbb{C}$ such that $J^2 = - I_{\mathsf{k}_\mathbb{C}}$. 
	Hence, in this case, $\mathcal{A}_x \cong \mathbb{C}$.
	This case occurs for $p-q \equiv_8 3, 7$ and $\mathsf{k}_\mathbb{C} = 2^{[\frac{n}{2}]}$.
	\medbreak
	{3) The {quaternionic} case}:
	\begin{equation}
		f_x = a_2 I_{\mathsf{k}_\mathbb{H}} + \displaystyle{\sum_{r=1}^{3}} b_r H_r, \ a_2, b_r \in \mathbb{R}, 
	\end{equation}
	where $H_r$ are matrices of order $\mathsf{k}_\mathbb{H} \times \mathsf{k}_\mathbb{H}$, $r=1, 2, 3$, such that $H_j H_k = - \delta_{jk}I_{\mathsf{k}_\mathbb{H}} + \displaystyle{\sum_{l=1}^{3}} \epsilon_{jkl} H_l$. In this case, $\mathcal{A}_x \cong \mathbb{H}$ and it occurs with the condition that $p-q \equiv_8 4, 5, 6$ for $\mathsf{k}_\mathbb{H} = 2^{[\frac{n}{2}]-1}$.
	\medbreak
	A bilinear mapping $B: \secp \times \secp \rightarrow \mathbb{R}$ is said to be admissible \cite{Bil} if the following three conditions hold:
	\medbreak
	a) $B$ is symmetric or skew-symmetric: $B(\alpha, \beta) = \sigma (B) B(\beta, \alpha)$, $\forall$ $\alpha, \beta \in \secp$, where $\sigma(B)=\pm 1$ is the symmetry of $B$.
	
	b) for $f \in \secextabrev$, the endomorphism $\lambda^\varGamma (f): \secp \rightarrow \secp$ is either $B$-symmetric or $B$-skew symmetric , this is, $B((\lambda^\varGamma (f))(\alpha), \beta) = \tau (B) B(\alpha, (\lambda^\varGamma (\widetilde{f}))(\beta))$, where $\tau(B)=\pm 1$ is the type of $B$.
	
	c) the splitting components $\varGamma (\mathcal{P}^+)$ and $\varGamma (\mathcal{P}^-)$ (when they exist) can be either 
	i) orthogonal: $B(\varGamma (\mathcal{P}^+), \varGamma (\mathcal{P}^-))=0$; or 
	ii) isotropic: $B(\varGamma (\mathcal{P}^+), \varGamma (\mathcal{P}^+))=0=B(\varGamma (\mathcal{P}^-), \varGamma (\mathcal{P}^-))$.\\
	In the first case the isotropy of $B$ is $i(B) =+ 1$ and in ii) it is $i(B) =- 1$.
	
	From the item $b)$ it is possible to define a  transpose of $\lambda^\varGamma (f)$ as a sort of adjoint operator: 
	\begin{equation}
		(\lambda^\varGamma (f))^\mathsf{T} = \tau (B) (\lambda^\varGamma (\widetilde{f})), \ \forall f \in \secextabrev.
	\end{equation}
	Thereafter, on a local coframe this transpose is given by
	\begin{equation}
		(\lambda^{\varGamma, I_k})^\mathsf{T} = (\tau (B))^{k} (-1)^{\frac{k(k-1)}{2}} \lambda^{\varGamma, I_k}.
	\end{equation} \index{Transpose}
	
	The values of $\sigma(B)$ and $\tau(B)$ are given onto fibers in Refs. \cite{bab1,type}, according to the following tables:
	\begin{table}[H]
		\begin{minipage}{.5\textwidth}
			\centering
			\begin{tabular}{c||c||c||c}
				\hline\hline
				$\mathcal{A}_x \cong$ & $p-q \equiv_8$ & $n \equiv_8$ & $\sigma(B)$\\
				\hline\hline
				\multirow{4}{*}{$\mathbb{R}$} 
				& \multirow{2}{*}{0, 2} & 0, 2 & 1  \\
				\hhline{~~--}
				&  & 4, 6 & $-1$  \\
				\hhline{~---}
				& \multirow{2}{*}{1} & 1, 7 & 1  \\
				\hhline{~~--}
				&  & 3, 5 & $-1$  \\
				\hline\hline
				\multirow{2}{*}{$\mathbb{C}$} 
				& \multirow{2}{*}{3, 7} & 1, 7 & 1  \\
				\hhline{~~--}
				&  & 3, 5 & $-1$  \\
				\hline\hline
				\multirow{4}{*}{$\mathbb{H}$} 
				& \multirow{2}{*}{4, 6} & 0, 2 & $-1$  \\
				\hhline{~~--}
				&  & 4, 6 & 1  \\
				\hhline{~---}
				& \multirow{2}{*}{5} & 1, 7 & $-1$  \\
				\hhline{~~--}
				&  & 3, 5 & 1  \\
				\hline\hline
			\end{tabular}
			\caption{\footnotesize Symmetry values of $B$.}
			\label{symmetry}
			
		\end{minipage} 
		\begin{minipage}{.5\textwidth}
			\centering
			\begin{tabular}{c||c||c||c}
				\hline\hline
				$\mathcal{A}_x \cong$ & $p-q \equiv_8$ & $n \equiv_8$ & $\tau(B)$\\
				\hline\hline
				\multirow{3}{*}{$\mathbb{R}$} & 0, 2 & & 1 \\
				\hhline{~---}
				& \multirow{2}{*}{1} & 1, 5 & 1  \\
				\hhline{~~--}
				&  & 3, 7 & $-1$  \\
				\hline\hline
				$\mathbb{C}$ & 3, 7 & & $-1$\\
				\hline\hline
				\multirow{3}{*}{$\mathbb{H}$} & 4, 6 & & 1 \\
				\hhline{~---}
				& \multirow{2}{*}{5} & 1, 5 & 1  \\
				\hhline{~~--}
				&  & 3, 7 & $-1$  \\
				\hline\hline
			\end{tabular}
			\caption{\footnotesize Type values of $B$.}
			\label{type}
		\end{minipage}
	\end{table}
	
	In what follows, consider the applications $\rho : \secp  \rightarrow  (\secp)^*$ such that $\rho (\alpha) (\beta) = B(\beta, \alpha) \in \mathbb{R}$ for $\alpha, \beta \in \secp$ and $\eta : \secp \times (\secp)^*  \rightarrow  \operatorname{End}(\secp)$ defined by $\eta (\alpha \otimes A) (\beta) = A(\beta) \alpha$, where $A: \secp \rightarrow \mathbb{R}$. Therefore the following mapping  can be defined:
	\begin{equation}
		E: = \eta \circ (Id_{\secp} \otimes \rho) : \secp \otimes \secp \rightarrow \operatorname{End}(\secp),
	\end{equation}
	which implies that the endomorphism $E_{\alpha,\beta}: = E(\alpha \otimes \beta): \secp \rightarrow \secp$ is such that
	\begin{equation}
		\begin{array}{rcl}
			E_{\alpha,\beta} (\gamma) &=& \eta ((Id_{\secp} \otimes \rho) (\alpha \otimes \beta)) (\gamma)\\
			&=& \eta (\alpha \otimes \rho(\beta))(\gamma)= (\rho(\beta)(\gamma)) (\alpha)\\
			&=& B(\gamma, \beta) \alpha, \quad \forall \gamma \in \secp.
		\end{array}
	\end{equation}
	Besides, for $\alpha_{1}, \alpha_2, \beta_1, \beta_2, \gamma \in \secp$, the composition of two of these endomorphisms results in a conformal endomorphism, with a conformal factor given by the smooth function $B$, 
	\begin{equation}
		\begin{array}{rcl}
			(E_{\alpha_{1},\beta_1} \circ E_{\alpha_2,\beta_2})(\gamma) &=& E_{\alpha_1,\beta_1}( E_{\alpha_2,\beta_2}(\gamma)) = E_{\alpha_1,\beta_1}( B(\gamma, \beta_2)\alpha_2)\\
			&=& B(\gamma, \beta_2) E_{\alpha_1,\beta_1}
			( \alpha_2) 
			\\
			&=& B(\alpha_2, \beta_1) (B(\gamma, \beta_2) \alpha_1)\\
			&=& B(\alpha_2, \beta_1) E_{\alpha_1,\beta_2}( \gamma).
		\end{array}
	\end{equation}
	Hence, the fundamental identity is established
	\begin{equation} \label{base}
		E_{\alpha_{1},\beta_1} \circ E_{\alpha_2,\beta_2} = B(\alpha_2, \beta_1) E_{\alpha_1,\beta_2},
	\end{equation}
	grounding the Fierz identities.
	
	In the normal case, the algebras $\mathcal{A}_x$ are isomorphic to $\mathbb{R}$. Hence, $\lambda^\varGamma$ is a surjective mapping, yielding  a global inverse mapping $(\lambda^\varGamma)^{-1}: \operatorname{End}_\mathbb{R} (\secp) = \operatorname{End} (\secp) \rightarrow \secextabrev$. In turn, if $\mathcal{A}_x \cong \mathbb{C}$ thus there is a partial inverse of $\lambda^\varGamma$ when it is restricted to endomorphisms over $\mathbb{C}$, i. e.,  $(\lambda_\mathbb{C}^\varGamma)^{-1}:=({{\lambda^\varGamma})^{-1}\big|}_{\operatorname{End}_\mathbb{C} (\secp)}: \operatorname{End}_\mathbb{C} (\secp) \rightarrow \secextabrev$. Similarly, in the quaternionic case the partial inverse of $\lambda^\varGamma$ is given by $(\lambda_\mathbb{H}^\varGamma)^{-1}:=({{\lambda^\varGamma})^{-1}\big|}_{\operatorname{End}_\mathbb{H} (\secp)}: \operatorname{End}_\mathbb{H} (\secp) \rightarrow \secextabrev$.
	
	When $\mathcal{A}_x \cong \mathbb{R}$ ($p-q \equiv_{8} 0, 1, 2$), no additional structure is needed, since $\lambda^\varGamma$ is invertible \cite{bab1,oku1}. If $\mathcal{A}_x \cong \mathbb{C}$ ($p-q \equiv_{8} 3, 7$) there exists an endomorphism $J \in \operatorname{Cen}_{\operatorname{End} (\secp)} (\lambda^\varGamma(\secextabrev)$ such that $J^2 = -Id_{\secp}$, in addition there is another endomorphism $D \in \operatorname{End} (\secp)$ so that $D^2 = (-1)^{\frac{p-q+1}{4}} Id_{\secp}$, $J \circ D = - D \circ J$ and $D \circ \lambda^\varGamma(f) =  \lambda^\varGamma(\#(f)) \circ D$, for all $f \in \secextabrev$. An example of this sort of structure is $J:= \pm \lambda^\varGamma(\textbf{v})$. In fact,
	\begin{equation}
		-Id_{\secp} = \lambda^\varGamma (- \unsecex) \stackrel{\text{Eq.}\eqref{volvol}}{=} \lambda^\varGamma (\textbf{v} \diamond \textbf{v}) = \lambda^\varGamma (\textbf{v}) \circ \lambda^\varGamma (\textbf{v}) = (\pm \lambda^\varGamma (\textbf{v}))^2,
	\end{equation}
	\begin{equation}
		\pm \lambda^\varGamma (\textbf{v}) \circ \lambda^\varGamma (f) = \pm \lambda^\varGamma (\textbf{v} \diamond f) = \pm \lambda^\varGamma (f \diamond \textbf{v}) = \pm \lambda^\varGamma (f) \circ \lambda^\varGamma (\textbf{v}).
	\end{equation}
	If $\mathcal{A}_x \cong \mathbb{H}$ ($p-q \equiv_{8} 4, 5, 6$), the structure in $\operatorname{End}(\secp)$ is given by $H_0= Id_{\secp}$, $H_1$, $H_2$ and $H_3$ in $\operatorname{Cen}_{\operatorname{End} (\secp)} (\lambda^\varGamma(\secextabrev)$ such that $H_i^2 = - Id_{\secp}$ and $H_j \circ H_k = - \delta_{jk} H_0 + \sum_{l=1}^{3} \epsilon_{jkl} H_l$.
	
	For each one of the isomorphism type of $\mathcal{A}_x \cong \mathbb{F}$ and $\alpha, \beta \in \secp$, let $E^{\mathbb{F}}_{\alpha, \beta}$ be the endomorphism $E_{\alpha, \beta}$ on the respective division algebra $\mathbb{F}$. The endomorphism $E^{\mathbb{F}}_{\alpha, \beta}$ can be locally  written as \cite{bab2}
	\begin{equation} \label{normal}
		E^{\mathbb{R}}_{\alpha, \beta} = \frac{\mathsf{k}_\mathbb{R}}{2^n} \sum_{I} (\tau(B))^{|I|} B(\alpha, \lambda^{\varGamma, I} (\beta)) \lambda^{\varGamma, I}
	\end{equation}
	for the \textit{normal case}, 
	\begin{eqnarray}
		\!\!\!\!\!\!\!\!\!\!\!\!\!E^{\mathbb{C}}_{\alpha, \beta} \!=\! \frac{\mathsf{k}_\mathbb{C}}{2^n} \sum_{I} (-1)^{|I|} B(\alpha, \lambda^{\varGamma, I} (\beta)) \lambda^{\varGamma, I}+ \frac{\mathsf{k}_\mathbb{C}}{2^n} (-1)^{\frac{p-q+1}{4}} \sum_{I}  B(\alpha, (D \circ \lambda^{\varGamma, I}) (\beta)) D \circ \lambda^{\varGamma, I} 
	\end{eqnarray}
	for the \textit{almost complex case} and
	\begin{equation}
		E^{\mathbb{H}}_{\alpha, \beta} = \frac{\mathsf{k}_\mathbb{H}}{2^n}  \sum_{i=0}^3 \sum_{I} (\tau(B))^{|I|} B(\alpha, (H_i \circ \lambda^{\varGamma, I}) (\beta)) H_i \circ \lambda^{\varGamma, I}
	\end{equation}
	for the \textit{quaternionic case}, where $I$ is an ordered index set and $B$ is an admissible bilinear mapping.
	
	To simplify the notation, the notations  $\mathrm{E}_{\alpha, \beta} := E^{\mathbb{R}}_{\alpha, \beta} $, $\textsf{E}_{\alpha, \beta} := E^{\mathbb{C}}_{\alpha, \beta} $ and $\mathbb{E}_{\alpha, \beta} := E^{\mathbb{H}}_{\alpha, \beta}$ shall be hereon adopted. Then, writing $\textsf{E}^{(0)}_{\alpha, \beta} = \frac{\mathsf{k}_\mathbb{C}}{2^n}  \sum_{I} (-1)^{|I|} B(\alpha, \lambda^{\varGamma, I} (\beta)) \lambda^{\varGamma, I}$ and $\textsf{E}^{(1)}_{\alpha, \beta} = \frac{\mathsf{k}_\mathbb{C}}{2^n} (-1)^{\frac{p-q+1}{4}} \sum_{I}  B(\alpha, (D \circ \lambda^{\varGamma, I}) (\beta)) \lambda^{\varGamma, I}$ yields 
	\begin{equation}\label{complex}
		\textsf{E}_{\alpha, \beta} = \textsf{E}^{(0)}_{\alpha, \beta} + D \circ \textsf{E}^{(1)}_{\alpha, \beta}.
	\end{equation}
	On the other hand, writing 
	\begin{equation} \label{quatern}
		\mathbb{E}^{(i)}_{\alpha, \beta} = \frac{\mathsf{k}_\mathbb{H}}{2^n} \sum_{I} (\tau(B))^{|I|} B(\alpha, (H_i \circ \lambda^{\varGamma, I}) (\beta)) \lambda^{\varGamma, I},
	\end{equation} 
	then
	\begin{equation}
		\mathbb{E}_{\alpha, \beta} = \sum_{i=0}^3 H_i \circ \mathbb{E}^{(i)}_{\alpha, \beta},
	\end{equation}
	since $H_0 = Id_{\secp}$.
	
	As the inverse of $\lambda^\varGamma$ was defined into the three cases, consider the following forms in $\secextabrev$:
	\begin{equation} \label{inv}
		\bar{\mathrm{E}}_{\alpha, \beta} := (\lambda_\mathbb{R}^\varGamma)^{-1} (\mathrm{E}_{\alpha, \beta}), \quad \bar{\textsf{E}}_{\alpha, \beta} := (\lambda_\mathbb{C}^\varGamma)^{-1} (\textsf{E}_{\alpha, \beta}), \quad \bar{\mathbb{E}}_{\alpha, \beta} := (\lambda_\mathbb{H}^\varGamma)^{-1} (\mathbb{E}_{\alpha, \beta}).
	\end{equation}
	From Eqs. \eqref{normal}, \eqref{complex} and \eqref{quatern}, their expansions read 
	\begin{subequations}
		\begin{eqnarray} \label{comp}
			\bar{\mathrm{E}}_{\alpha, \beta} &=& \frac{\mathsf{k}_\mathbb{R}}{2^n} \sum_{I} (\tau(B))^{|I|} B(\alpha, \lambda^{\varGamma, I} (\beta)) e^I,\\
			\bar{\textsf{E}}^{(0)}_{\alpha, \beta} &=& \frac{\mathsf{k}_\mathbb{C}}{2^n}  \sum_{I} (-1)^{|I|} B(\alpha, \lambda^{\varGamma, I} (\beta)) e^I, \\
			\bar{\textsf{E}}^{(1)}_{\alpha, \beta} &=& \frac{\mathsf{k}_\mathbb{C}}{2^n} (-1)^{\frac{p-q+1}{4}} \sum_{I}  B(\alpha, (D \circ \lambda^{\varGamma, I}) (\beta)) e^I, \\
			\bar{\mathbb{E}}^{(0)}_{\alpha, \beta} &=& \frac{\mathsf{k}_\mathbb{H}}{2^n} \sum_{I} (\tau(B))^{|I|} B(\alpha,  \lambda^{\varGamma, I} (\beta)) e^I\\
			\text{and} \quad \bar{\mathbb{E}}^{(i)}_{\alpha, \beta} &=& \frac{\mathsf{k}_\mathbb{H}}{2^n} \sum_{I} (\tau(B))^{|I|} B(\alpha, (H_i \circ \lambda^{\varGamma, I}) (\beta)) e^I, \ \text{for} \ i=1, 2, 3.
		\end{eqnarray}
	\end{subequations}
	Locally, for an index set $I_k$, they are respectively expressed as:
	\begin{subequations}
		\begin{eqnarray} \label{compkr}
			\bar{\mathrm{E}}_{\alpha, \beta} &=& \frac{\mathsf{k}_\mathbb{R}}{2^n} \sum_{\substack{k=0 \\ I_k=\text{ordered}}}^n \frac{1}{k!}(\tau(B))^{k} B(\alpha, \lambda^{\varGamma, I_k} (\beta)) e^{I_k}\\\label{compkc1}
			\bar{\textsf{E}}^{(0)}_{\alpha, \beta} &=& \frac{\mathsf{k}_\mathbb{C}}{2^n}  \sum_{\substack{k=0 \\ I_k=\text{ordered}}}^n \frac{1}{k!} (-1)^{k} B(\alpha, \lambda^{\varGamma, I_k} (\beta)) e^{I_k} \\\label{compkc2}
			\bar{\textsf{E}}^{(1)}_{\alpha, \beta} &=& \frac{\mathsf{k}_\mathbb{C}}{2^n} (-1)^{\frac{p-q+1}{4}} \sum_{\substack{k=0 \\ I_k=\text{ordered}}}^n \frac{1}{k!}  (-1)^k B(\alpha, (D \circ \lambda^{\varGamma, I_k}) (\beta)) e^{I_k} \\
			\bar{\mathbb{E}}^{(0)}_{\alpha, \beta} &=& \frac{\mathsf{k}_\mathbb{H}}{2^n} \sum_{\substack{k=0 \\ I_k=\text{ordered}}}^n \frac{1}{k!} (\tau(B))^{k} B(\alpha,  \lambda^{\varGamma, I_k} (\beta)) e^{I_k}\\
			\bar{\mathbb{E}}^{(i)}_{\alpha, \beta} &=& \frac{\mathsf{k}_\mathbb{H}}{2^n} \sum_{\substack{k=0 \\ I_k=\text{ordered}}}^n \frac{1}{k!} (\tau(B))^{k} B(\alpha, (H_i \circ \lambda^{\varGamma, I_k}) (\beta)) e^{I_k}, \ \text{for} \ i=1, 2, 3.
		\end{eqnarray}
	\end{subequations}
	The Fierz identities come from Eq. \eqref{base}, being  written according to each $\mathbb{F}$:
	\begin{equation}
		\bar{X}_{\alpha_{1},\beta_1} \diamond \bar{X}_{\alpha_2,\beta_2} = B(\alpha_2, \beta_1) \bar{X}_{\alpha_1,\beta_2}, \ \forall \alpha_{1}, \alpha_{2}, \beta_1, \beta_2 \in \secp,
	\end{equation}
	where $X= \mathrm{E}$, $\textsf{E}$ or $\mathbb{E}$.
	Therefore, using the components defined in Eq. \eqref{comp}, the Fierz identities in the normal case read \cite{bab2,oku2}:
	\begin{eqnarray}\label{finc}
		&\bar{\mathrm{E}}_{\alpha_{1},\beta_1} \diamond \bar{\mathrm{E}}_{\alpha_2,\beta_2} = B(\alpha_2, \beta_1) \bar{\mathrm{E}}_{\alpha_1,\beta_2},&
	\end{eqnarray}  
	Besides, in the almost complex case we have
	\begin{eqnarray}\label{fiacc}
		i) \ \bar{\textsf{E}}^{(0)}_{\alpha_{1},\beta_1} \diamond \bar{\textsf{E}}^{(0)}_{\alpha_2,\beta_2} + (-1)^{\frac{p-q+1}{4}} \#(\bar{\textsf{E}}^{(1)}_{\alpha_{1},\beta_1}) \diamond \bar{\textsf{E}}^{(1)}_{\alpha_2,\beta_2}&=& B(\alpha_2, \beta_1) \bar{\textsf{E}}^{(0)}_{\alpha_1,\beta_2}\\
		ii) \ \#(\bar{\textsf{E}}^{(0)}_{\alpha_{1},\beta_1}) \diamond \bar{\textsf{E}}^{(1)}_{\alpha_2,\beta_2} +  \bar{\textsf{E}}^{(1)}_{\alpha_{1},\beta_1} \diamond \bar{\textsf{E}}^{(0)}_{\alpha_2,\beta_2}&=& B(\alpha_2, \beta_1) \bar{\textsf{E}}^{(1)}_{\alpha_1,\beta_2},
	\end{eqnarray} 
	\noindent whereas the quaternionic case, for $i=1, 2,3$, reads
	\begin{eqnarray}\label{fiqc}
		&	i) \ \bar{\mathbb{E}}^{(0)}_{\alpha_{1},\beta_1} \diamond \bar{\mathbb{E}}^{(0)}_{\alpha_2,\beta_2} - \sum_{i=1}^{3} (\bar{\mathbb{E}}^{(i)}_{\alpha_{1},\beta_1} \diamond \bar{\mathbb{E}}^{(i)}_{\alpha_2,\beta_2})= B(\alpha_2, \beta_1) \bar{\mathbb{E}}^{(0)}_{\alpha_1,\beta_2}\\&
		ii) \ {\mathbb{E}}^{(0)}_{\alpha_{1},\beta_1}\!\diamond\!\bar{\mathbb{E}}^{(i)}_{\alpha_2,\beta_2}\!+\!{\mathbb{E}}^{(i)}_{\alpha_{1},\beta_1}\!\diamond\!\bar{\mathbb{E}}^{(0)}_{\alpha_2,\beta_2}\!+\!\sum_{i=1}^{3} (\epsilon_{ijk} \bar{\mathbb{E}}^{(j)}_{\alpha_{1},\beta_1}\!\diamond\!\bar{\mathbb{E}}^{(k)}_{\alpha_2,\beta_2})\!=\!B(\alpha_2, \beta_1) \bar{\mathbb{E}}^{(i)}_{\alpha_1,\beta_2}.&
	\end{eqnarray}

	\section{Spinor classes on signature (1, 2)}
	
	New classes of spinors have been found under the formalism of the Graf product and the Fierz identities, for signatures $(7, 0)$, $(6,1)$, and $(1, 4)$, respectively in Refs. \cite{Bon14,Bonora:2015ppa,brito}. In this section we will find {\color{black}{three non-trivial}} new classes of spinor fields on 3-dimensional pseudo-Riemannian manifolds with signature $(1, 2)$, employing this method. {\color{black}{One of the main motivations to explore new classes of spinor fields in such dimension and signature is the existence of anyons in quantum field theory, interpolating between  Fermi--Dirac and Bose--Einstein statistics \cite{Frohlich:1988qh}.  Besides the formal aspects, one can find a vast variety of applications regarding $(1,2)$ spinor fields, playing a prominent role on the  stage of condensed matter physics. Fermions quantum fields constructed upon these spinors  can  describe superconductors \cite{Grignani:1995vn} and semimetals, with particular attention to the graphene \cite{Gavrilov:2012jk,Gonzalez:2009je,Dutreix:2013jva}.}}
	
	In spaces of signature $(1,2)$, namely $n=3$, we have $p-q\equiv_8 7$. Hence, the main subalgebra is given by  $\mathcal{A}_x \cong \mathbb{C}$. Now, given an element, $\alpha \in \secp$ define 
	\begin{equation}
		\alpha_\pm := \frac{1}{2}(\alpha \pm D(\alpha)).
	\end{equation}
	If $\alpha, \beta \in \secp$ and $f \in \secextabrev$, then the almost complex case yields \cite{bab1} 
	\begin{equation}\label{b}
		B(\alpha, \lambda^\varGamma(f) (\beta)) = \left\{ \begin{array}{l}
			B(\alpha_+, \lambda^\varGamma(f) (\beta_+)) - (-1)^\frac{n(n+1)}{2} B(\alpha_-, \lambda^\varGamma(f) (\beta_-)),  \quad \text{if} \ f \in \varGamma^+\\
			B(\alpha_+, J \circ \lambda^\varGamma(f) (\beta_-)) + (-1)^\frac{n(n+1)}{2} B(\alpha_-, J \circ \lambda^\varGamma(f) (\beta_+)), \quad \text{if} \ f \in \varGamma^-
		\end{array} \right. .
	\end{equation}
	Since $\secpm = {\textstyle \frac{1}{2}}(Id_{\secp} \pm D)\secp$, whenever $\alpha \in \secpe$, then there exists an element  $\beta \in \secp$ such that $\alpha = {\textstyle \frac{1}{2}}(\beta + D(\beta))$, yielding 
	\begin{eqnarray}\nonumber
		\alpha_+ &=& {\textstyle \frac{1}{2}}( {\textstyle \frac{1}{2}}(\beta + D(\beta)) + D( {\textstyle \frac{1}{2}}(\beta + D(\beta))))\\\nonumber
		&=&  {\textstyle \frac{1}{2}}( {\textstyle \frac{1}{2}}\beta + D(\beta)  +{\textstyle \frac{1}{2}} (-1)^{\frac{p-q+1}{4}}\beta)\\
		&=&  {\textstyle \frac{1}{2}}( {\textstyle \frac{1}{2}}\beta + D(\beta)  +{\textstyle \frac{1}{2}} (-1)^{\frac{1-2+1}{4}}\beta)= \alpha
	\end{eqnarray}
	and 
	\begin{eqnarray}\nonumber
		\alpha_- &=& {\textstyle \frac{1}{2}}( {\textstyle \frac{1}{2}}(\beta + D(\beta)) - D( {\textstyle \frac{1}{2}}(\beta + D(\beta))))\\
		&=&  {\textstyle \frac{1}{2}}( {\textstyle \frac{1}{2}}\beta + {\textstyle \frac{1}{2}} D(\beta) - {\textstyle \frac{1}{2}} D(\beta) -{\textstyle \frac{1}{2}} \beta)=0.
	\end{eqnarray}
	Hence, by Eq. \eqref{b} for $\alpha \in \secpe$ to hold,  $B(\alpha, \lambda^{\varGamma, I_k} (\alpha))=0$, whenever $k$ is odd. Then,  this result applied on Eq. \eqref{compkc1} implies that
	\begin{eqnarray}\nonumber
		\bar{\textsf{E}}^{(0)}_{\alpha, \alpha} &=& \frac{\mathsf{k}_\mathbb{C}}{2^n}  \sum_{\substack{k=0 \\ I_k=\text{ordered}}}^n \frac{1}{k!} (-1)^{k} B(\alpha, \lambda^{\varGamma, I_k} (\alpha)) e^{I_k} 
		= \frac{\mathsf{k}_\mathbb{C}}{2^n}  \sum_{\substack{k=\text{even} \\ I_k=\text{ordered}}}^n \frac{1}{k!} (-1)^{k} B(\alpha, \lambda^{\varGamma, I_k} (\alpha)) e^{I_k}\\\nonumber
		&=& \frac{1}{4}  \sum_{\substack{k=\text{even} \\ I_k=\text{ordered}}}^3 \frac{1}{k!} B(\alpha, \lambda^{\varGamma, I_k} (\alpha)) e^{I_k}.
	\end{eqnarray}
	
	On the other hand,  if $\alpha, \beta \in \secp$ and $f \in \secextabrev$, thus \cite{bab1}
	\begin{equation}\label{bd}
		B(\alpha, D \circ \lambda^\varGamma(f) (\beta)) \!=\! \left\{ \begin{array}{l}
			B(\alpha_+, \lambda^\varGamma(f) (\beta_+)) \!-\! (-1)^\frac{n(n+1)}{2} B(\alpha_-, \lambda^\varGamma(f) (\beta_-)),  \;\;\; \text{if} \ f \in \varGamma^+\\
			B(\alpha_+, J \circ \lambda^\varGamma(f) (\beta_-)) \!+\! (-1)^\frac{n(n+1)}{2} B(\alpha_-, J \circ \lambda^\varGamma(f) (\beta_+)), \;\;\; \text{if} \ f \in \varGamma^-
		\end{array} \right. .
	\end{equation} 
	Then, for $\alpha \in \secpe$, it follows that 
	\begin{equation}\label{bde}
		B(\alpha, D \circ \lambda^{\varGamma, I_k} (\alpha)) = \left\{ \begin{array}{l}
			B(\alpha, \lambda^{\varGamma, I_k} (\alpha)),  \quad \text{if} \ k \ \text{is even}\\
			0, \quad \text{if} \ k \ \text{is odd}
		\end{array} \right. .
	\end{equation}
	
	Considering Eq. \eqref{compkc2} and element $\alpha \in \secpe$, it yields
	\begin{eqnarray}\nonumber
		\bar{\textsf{E}}^{(1)}_{\alpha, \alpha} &=& \frac{\mathsf{k}_\mathbb{C}}{2^n} (-1)^{\frac{p-q+1}{4}} \sum_{\substack{k=0 \\\nonumber I_k=\text{ordered}}}^n \frac{1}{k!} (-1)^k B(\alpha, (D \circ \lambda^{\varGamma, I_k}) (\alpha)) e^{I_k} \\\nonumber
		&=& \frac{1}{4} (-1)^{\frac{1-2+1}{4}} \sum_{\substack{k=\text{even} \\ I_k=\text{ordered}}}^3 \frac{1}{k!}  (-1)^{\text{even}} B(\alpha,  \lambda^{\varGamma, I_k} (\alpha)) e^{I_k} = \bar{\textsf{E}}^{(0)}_{\alpha, \alpha}\\\nonumber
		&=& \frac{1}{4} (B(\alpha, Id_{\secp} (\alpha)) \unsecex + \sum_{\substack{I_2=\text{ordered}}} \frac{1}{2!} B(\alpha,  \lambda^{\varGamma, I_2} (\alpha)) e^{I_2})\\
		&=& \frac{1}{4} (B(\alpha, \alpha) \unsecex + \frac{1}{2} \sum_{i, j=1}^3  B(\alpha,  \lambda^\varGamma(e^i) \circ \lambda^\varGamma(e^i) (\alpha)) e^i \wedge e^j).
	\end{eqnarray}
	In addition, on the almost complex case we have $\tau(B)=-1$, then
	\begin{equation}
		(\lambda^{\varGamma, I_k})^\mathsf{T} = (-1)^{k} (-1)^{\frac{k(k-1)}{2}} \lambda^{\varGamma, I_k} = (-1)^{\frac{k(k+1)}{2}} \lambda^{\varGamma, I_k},
	\end{equation}
	for $k=1, 2, 3.$
	
	Since $B$ is skew-symmetric ($\sigma(B)=-1$), it implies that  $(\lambda^{\varGamma, I_k})^\mathsf{T}$ is equal to $ - \lambda^{\varGamma, I_k}$, this implies that
	\begin{equation}
		(-1)^{\frac{k(k+1)}{2}} = -1,
	\end{equation}
	whenever $k(k+1)$ is divisible by 2 only once. Then, it follows that $k=1, 2, 5, 6$, otherwise $\lambda^{\varGamma, I_k} = 0$. Besides, for $k \neq 0$, the relation $\lambda^{\varGamma, I_0} = Id_{\secp}$ holds.
	
	Hence, defining the bilinear covariants 
	\begin{equation}
		\phi_0 := B(\alpha, \alpha) \unsecex \qquad \ \ \text{and} \qquad \ \ \phi_2 := \frac{1}{2} \sum_{i, j=1}^3  B(\alpha,  \lambda^\varGamma(e^i) \circ \lambda^\varGamma(e^j) (\alpha)) e^i \wedge e^j,
	\end{equation}
	therefore 
	\begin{equation}
		\bar{\textsf{E}}^{(0)}_{\alpha, \alpha} = \bar{\textsf{E}}^{(1)}_{\alpha, \alpha} = \frac{1}{4} (\phi_0 + \phi_2).
	\end{equation}
	
	Thus, the two Fierz identities for the almost complex case \eqref{fiacc} consist of the same identity:
	\begin{equation}
		\frac{1}{16} (\phi_0\!+\!\phi_2) \diamond (\phi_0\!+\!\phi_2) + \frac{1}{16} (-1)^{\frac{1-2+1}{2}} (\phi_0\!+\!\phi_2) \diamond (\phi_0\!+\!\phi_2) = \frac{1}{4} B(\alpha, \alpha) (\phi_0\!+\!\phi_2)
	\end{equation}
	which implies that
	\begin{eqnarray}\label{12} \nonumber
		\frac{1}{8}(\phi_0 \diamond \phi_0 + \phi_2 \diamond \phi_0 +\phi_0 \diamond \phi_2 +\phi_2 \diamond \phi_2)= \frac{1}{4} B(\alpha, \alpha) (\phi_0+\phi_2)\\ 
		\Rightarrow  \frac{1}{8}(\underbrace{\phi_0 \wedge \phi_0}_{0\text{-form}} + \underbrace{\phi_2 \wedge \phi_0}_{2\text{-form}} +\underbrace{\phi_0 \wedge \phi_2}_{2\text{-form}} +\phi_2 \diamond \phi_2)= \frac{1}{4} B(\alpha, \alpha) (\phi_0+\phi_2)
	\end{eqnarray}
	where
	\begin{eqnarray}
		\phi_2 \diamond \phi_2 &=& \displaystyle{\sum^{2}_{r=0}} \dfrac{(-1)^{r(2-r)+[\frac{r}{2}]}}{r!} \phi_2 \wedge_r \phi_2=  \underbrace{\phi_2 \wedge \phi_2}_{=0} - \underbrace{\phi_2 \wedge_1 \phi_2}_{2\text{-form}} - \frac{1}{2} \underbrace{\phi_2 \wedge_2 \phi_2}_{0\text{-form}}.
	\end{eqnarray}
	
	Hence, Eq. \eqref{12} becomes
	\begin{equation}\label{12eq}
		\phi_0 \wedge \phi_0 + \phi_2 \wedge \phi_0 +\phi_0 \wedge \phi_2 - \phi_2 \wedge_1 \phi_2 - \frac{1}{2} \phi_2 \wedge_2 \phi_2 = 2 B(\alpha, \alpha) \phi_0+2 B(\alpha, \alpha)\phi_2.
	\end{equation}
	Now, since $\phi_0 \wedge \phi_0 = B(\alpha, \alpha) \phi_0$ and $\phi_0 \wedge \phi_2 = B(\alpha, \alpha) \phi_2 = \phi_2 \wedge \phi_0$, regarding Eq. \eqref{12eq} and equaling the similar forms yields 
	\begin{eqnarray}
		B(\alpha, \alpha) \phi_0 - \frac{1}{2} \phi_2 \wedge_2 \phi_2 = 2 B(\alpha, \alpha) \phi_0, \\
		B(\alpha, \alpha) \phi_2 + B(\alpha, \alpha) \phi_2 -  \phi_2 \wedge_1 \phi_2 = 2 B(\alpha, \alpha)\phi_2.
	\end{eqnarray}
	
	Hence, the Fierz identities on the almost complex case for signature $(1, 2)$ and $\alpha \in \secpe$ {\color{black}{reduce to}}
	\begin{equation}\label{12eq1}
		{
			\phi_2 \wedge_2 \phi_2 = -2 B(\alpha, \alpha) \phi_0}
	\end{equation}
	\begin{equation}\label{12eq2}
		{
			\phi_2 \wedge_1 \phi_2=0.}
	\end{equation}
	
	In this way, new classes of Majorana spinors on a 3-dimensional manifold of signature $(1, 2)$ are given by the two forms $\phi_0$ and $\phi_2$ satisfying Eqs. \eqref{12eq1} and \eqref{12eq2} according the following four classes:
	\medbreak
	1) $\phi_0=0$, $\phi_2=0$,
	
	2) $\phi_0\neq0$, $\phi_2=0$,
	
	3) $\phi_0=0$, $\phi_2\neq0$, 
	
	4) $\phi_0\neq0$, $\phi_2\neq0$.
	
	\medbreak
	
	{\color{black}{The first above class is a trivial one. The spinor fields classes 2) and 4) have fermionic representatives in the literature (see, e. g., Refs. \cite{Gavrilov:2012jk,Gonzalez:2009je,Dutreix:2013jva}). However, up to our knowledge, the above class 3) has no identified representative, yet. Therefore, this new class can provide, via the reconstruction theorem, 
			new spinor fields that can be used in the construction of new quantum fields that potentially represent fermionic states in, for instance, condensed matter systems.}}

	In the next section, new classes of pinors on spaces of signature $(9,0)$ shall be explored.

	\section{Pinor classes on (9, 0)}
	\label{sec5}
	
	To analyse pinors on 9-dimensional pseudo-Riemannian manifolds of signature $(9,0)$, in this case $n=9$, $p-q\equiv_8 1$ and therefore the main subalgebra is given by  $\mathcal{A}_x \cong \mathbb{R}$.
	Since $\dim \mathcal{M}$ is odd and $p-q\equiv_8 1$, the truncated model shall be adopted to find new classes of pinors in the aforementioned spacetime dimension and signature. For this signature, the type $\tau (B)$ is equal to 1, implying that 
	\begin{equation}
		(\lambda^{\varGamma, I_k})^\mathsf{T} = (-1)^{\frac{k(k-1)}{2}} \lambda^{\varGamma, I_k}.
	\end{equation}
	The symmetry $\sigma(B)$ is equal to 1, implying that  $(-1)^{\frac{k(k-1)}{2}} = 1$. Hence, $k(k-1)$ is divisible by 4, and consequently $k=0, 1, 4, 5, 8, 9$.
	Using it into Eq. \eqref{compkr} for $\alpha \in \secp$ yields
	\begin{eqnarray}\nonumber
		\bar{\mathrm{E}}_{\alpha, \alpha} &=& \frac{\mathsf{k}_\mathbb{R}}{2^n} \sum_{\substack{k=0 \\ I_k=\text{ordered}}}^n \frac{1}{k!} (\tau(B))^{k} B(\alpha,  \lambda^{\varGamma, I_k} (\alpha)) e^{I_k}\\ 
		&=& \frac{1}{32} \sum_{\substack{k=0 \\ I_k=\text{ordered}}}^9 \frac{1}{k!} B(\alpha,  \lambda^{\varGamma, I_k} (\alpha)) e^{I_k}.
	\end{eqnarray}
	
	From the truncated model, new pinor classes will be found on the space
	\begin{equation}
		\varGamma_L = \bigoplus_{k=0}^{[\frac{9}{2}]} \secextkabrev = \bigoplus_{k=0}^{4} \secextkabrev.
	\end{equation}
	
	In this way, we write the truncated version of $\bar{\mathrm{E}}_{\alpha, \alpha}$ as
	\begin{equation}
		\bar{\mathrm{E}}_{\alpha, \alpha}^{L} := \frac{1}{32} \sum_{\substack{k=0 \\ I_k=\text{ordered}}}^4 \frac{1}{k!} B(\alpha,  \lambda^{\varGamma, I_k} (\alpha)) e^{I_k}
	\end{equation}
	
	Defining the following truncated forms 
	\begin{subequations}
		\begin{eqnarray}
			\psi_0 &=& B(\alpha, \alpha), \\
			\psi_1 &=& \sum_{i=1}^4 B(\alpha,  \lambda^\varGamma(e^i) (\alpha)) e^i,\\
			\psi_4 &=& \frac{1}{4!}\sum_{j,k, l, m=1}^4 B(\alpha,  \lambda^\varGamma(e^j)\!\circ\!\lambda^\varGamma(e^k)\!\circ\!\lambda^\varGamma(e^l)\!\circ\!\lambda^\varGamma(e^m) (\alpha)) e^j\!\wedge\!e^k\!\wedge\!e^l\!\wedge\!e^m,
		\end{eqnarray}
	\end{subequations}
	then
	\begin{equation}
		\bar{\mathrm{E}}_{\alpha, \alpha}^{L} = \frac{1}{32} (\psi_0 + \psi_1 + \psi_4) \in \varGamma_L.
	\end{equation}
	
	The truncated product $\blkdiam_+$ defined in Ref. \cite{frame} will be used for elements in $\varGamma_L$ and will be denoted by $\blkdiam$. Remembering that the Fierz identity for this case is given by 
	\begin{equation}\label{barE}
		\bar{\mathrm{E}}_{\alpha,\alpha} \diamond \bar{\mathrm{E}}_{\alpha,\alpha} = B(\alpha, \alpha) \bar{\mathrm{E}}_{\alpha,\alpha},
	\end{equation} then its truncated version regarding $\blkdiam$ is given by
	\begin{eqnarray}\nonumber
		\bar{\mathrm{E}}_{\alpha, \alpha}^{L} \ \blkdiam \  \bar{\mathrm{E}}_{\alpha, \alpha}^{L} &=& 2 P_L (P_+ (\bar{\mathrm{E}}_{\alpha, \alpha}^{L} \diamond \bar{\mathrm{E}}_{\alpha, \alpha}^{L}))= 2 P_L (P_+ (B(\alpha, \alpha) \bar{\mathrm{E}}_{\alpha, \alpha}^{L})) \\ \nonumber
		&=&  B(\alpha, \alpha) P_L ( \bar{\mathrm{E}}_{\alpha, \alpha}^{L} + \star (\bar{\mathrm{E}}_{\alpha, \alpha}^{L}))= B(\alpha, \alpha) P_L ( \bar{\mathrm{E}}_{\alpha, \alpha}^{L} + \bar{\mathrm{E}}_{\alpha, \alpha}^{L} \diamond \textbf{v})\\
		&=& B(\alpha, \alpha) ( \bar{\mathrm{E}}_{\alpha, \alpha}^{L} +  P_L (  \bar{\mathrm{E}}_{\alpha, \alpha}^{L} \diamond \textbf{v})).
	\end{eqnarray}
	In particular, Eq. \eqref{fvol} yields 
	\begin{eqnarray}\nonumber
		\bar{\mathrm{E}}_{\alpha, \alpha}^{L} \diamond \textbf{v} &=& \frac{1}{32} (\psi_0 \diamond \textbf{v} + \psi_1 \diamond \textbf{v} + \psi_4 \diamond \textbf{v})= \frac{1}{32} (\psi_0 \wedge \textbf{v} + (-1)^{[\frac{1}{2}]} \psi_1 \wedge_1 \textbf{v} + (-1)^{[\frac{4}{2}]} {\textstyle\frac{1}{4!}} \psi_4 \wedge_4 \textbf{v})\\
		&=& \frac{1}{32} (\underbrace{\psi_0 \wedge \textbf{v}}_{9-\text{form}} + \underbrace{\psi_1 \wedge_1 \textbf{v}}_{8-\text{form}} + {\textstyle\frac{1}{4!}} \underbrace{\psi_4 \wedge_4 \textbf{v}}_{5-\text{form}}).
	\end{eqnarray}
	It implies that $P_L(\bar{\mathrm{E}}_{\alpha, \alpha}^{L} \diamond \textbf{v})=0$, hence 
	\begin{equation}\label{EE}
		\bar{\mathrm{E}}_{\alpha, \alpha}^{L} \ \blkdiam \ \bar{\mathrm{E}}_{\alpha, \alpha}^{L} =B(\alpha, \alpha)  \bar{\mathrm{E}}_{\alpha, \alpha}^{L}.
	\end{equation}
	
	Considering Eq. \eqref{barE},  Eq. \eqref{EE} becomes
	\begin{eqnarray}
		&&	{\textstyle\frac{1}{32^2}} (\psi_0 + \psi_1 + \psi_4) \ \blkdiam \ (\psi_0 + \psi_1 + \psi_4) = B(\alpha, \alpha) {\textstyle\frac{1}{32}} (\psi_0 + \psi_1 + \psi_4)
	\end{eqnarray} or, equivalently, 
	\begin{eqnarray} \!\!\!\!\!\!\!&&\psi_0\,\blkdiam\,  \psi_0 \!+\! \psi_0\,\blkdiam\,  \psi_1 \!+\! \psi_0\,\blkdiam\,  \psi_4 \!+\! \psi_1\,\blkdiam\,  \psi_0 \!+\! \psi_1\,\blkdiam\,  \psi_1 \!+\! \psi_1\,\blkdiam\,  \psi_4 \!+\! \psi_4\,\blkdiam\,  \psi_0 \!+\! \psi_4\,\blkdiam\,  \psi_1 \!+\! \psi_4\,\blkdiam\,  \psi_4 \nonumber\\&&\qquad\qquad\qquad= 32  B(\alpha, \alpha) (\psi_0 + \psi_1 + \psi_4). \label{rc1}
	\end{eqnarray}
	The following expressions 
	\begin{subequations}
		\begin{eqnarray}
			\psi_0 \ \blkdiam \  \psi_0 &=&B(\alpha, \alpha)  \psi_0 \in \varGamma({{\textstyle \bigwedge}}^0),\label{eq123}\\
			\psi_0 \ \blkdiam \  \psi_1 &=&B(\alpha, \alpha)  \psi_1 \in \varGamma({{\textstyle \bigwedge}}^1),\label{eq1231}\\
			\psi_0 \ \blkdiam \  \psi_4&=&B(\alpha, \alpha)  \psi_4 \in \varGamma({{\textstyle \bigwedge}}^4),\\\label{eq1232}
			\psi_1 \ \blkdiam \  \psi_0 &=&  B(\alpha, \alpha) (\psi_1+ P_L(\psi_1 \diamond \textbf{v})) = B(\alpha, \alpha)  \psi_1 \in \varGamma({{\textstyle \bigwedge}}^1),\\\label{eq1233}
			\psi_1 \ \blkdiam \  \psi_1 &=&\psi_1 \wedge \psi_1 + \psi_1 \wedge_1 \psi_1 \in \varGamma({{\textstyle \bigwedge}}^2) \oplus \varGamma({{\textstyle \bigwedge}}^0),\\\label{eq1234}
			\psi_1 \ \blkdiam \  \psi_4 
			&=& \psi_1 \wedge_1 \psi_4 + \star(\psi_1 \wedge \psi_4) \in \varGamma({{\textstyle \bigwedge}}^3) \oplus \varGamma({{\textstyle \bigwedge}}^4),\\\label{eq1235}
			\psi_4 \ \blkdiam \  \psi_0 &=&  B(\alpha, \alpha)  \psi_4 \in \varGamma({{\textstyle \bigwedge}}^4),\\
			\psi_4 \ \blkdiam \  \psi_1&=& - \psi_1 \wedge_1 \psi_4 + \star(\psi_1 \wedge \psi_4) \in \varGamma({{\textstyle \bigwedge}}^3) \oplus \varGamma({{\textstyle \bigwedge}}^4),\label{eq1236}\\
			\psi_4 \ \blkdiam \  \psi_4 &=& -\!{\textstyle\frac{1}{2}} \psi_4 \wedge_2 \psi_4\!+\!{\textstyle\frac{1}{3!}} \psi_4 \wedge_3 \psi_4\!+\!{\textstyle\frac{1}{4!}} \psi_4 \wedge_4 \psi_4 + \star (\psi_4 \wedge \psi_4) - \star (\psi_4 \wedge_1 \psi_4)\\
			&& \in \varGamma({{\textstyle \bigwedge}}^4) \oplus \varGamma({{\textstyle \bigwedge}}^2) \oplus \varGamma({{\textstyle \bigwedge}}^0)\oplus \varGamma({{\textstyle \bigwedge}}^1)\oplus \varGamma({{\textstyle \bigwedge}}^3),\label{eq124}
		\end{eqnarray}
	\end{subequations}
	are explicitly computed in the Appendix, for the sake of conciseness. 
	
	Hence, Eq. \eqref{rc1} becomes
	\begin{eqnarray}\nonumber
		&&B(\alpha, \alpha)  \psi_0\!+ 2 B(\alpha, \alpha)  \psi_1 \!+ 2B(\alpha, \alpha)  \psi_4\!+\!\psi_1 \wedge \psi_1\!+\!\psi_1 \wedge_1 \psi_1\!+ 2\star(\psi_1\wedge\psi_4)\!-\!{\textstyle\frac{1}{2}} \psi_4 \wedge_2 \psi_4\\ 
		&&\!+\!{\textstyle\frac{1}{3!}} \psi_4 \wedge_3 \psi_4\!+\!{\textstyle\frac{1}{4!}} \psi_4 \wedge_4 \psi_4 \!+\! \star (\psi_4\wedge \psi_4)-\star (\psi_4 \wedge_1 \psi_4) = 32  B(\alpha, \alpha) (\psi_0\!+\!\psi_1\!+\!\psi_4). \label{rc4}
	\end{eqnarray}
	
	Equaling the similar forms in Eq. \eqref{rc4} it follows that the Fierz identities for the normal case on spaces of signature $(9, 0)$ are given by the five equations below:
	\begin{eqnarray}\label{rcfi1}
		\psi_1 \wedge_1 \psi_1 +{\textstyle\frac{1}{4!}} \psi_4 \wedge_4 \psi_4 &=& 31 B(\alpha, \alpha)  \psi_0,
		\\\label{rcfi2}
		\star (\psi_4\wedge \psi_4) &=& 30 B(\alpha, \alpha)  \psi_1,
		\\\label{rcfi3}
		\psi_1 \wedge \psi_1 + {\textstyle\frac{1}{3!}} \psi_4 \wedge_3 \psi_4 &=&0,\\\label{rcfi4}
		\star (\psi_4 \wedge_1 \psi_4) &=&0,\\
		\label{rcfi5}
		\star(4 \psi_1\wedge\psi_4) -\psi_4 \wedge_2 \psi_4 &=& 60B(\alpha, \alpha)  \psi_4.
	\end{eqnarray}
	
	Therefore, pinors on a 9-dimensional manifold of signature $(9, 0)$ are given by the forms $\psi_0$, $\psi_1$ and $\psi_4$ satisfying Eqs. \eqref{rcfi1} to \eqref{rcfi5} according the following 8 classes:
	\medbreak
	1) $\psi_0=0$, $\psi_1\neq0$, $\psi_4\neq0$,
	
	2) $\psi_1=0$, $\psi_0\neq0$, $\psi_4\neq0$,
	
	3) $\psi_4=0$, $\psi_0\neq0$, $\psi_1\neq0$,
	
	4) $\psi_0=\psi_1=0$, $\psi_4\neq0$,
	
	5) $\psi_0=\psi_4=0$, $\psi_1\neq0$,
	
	6) $\psi_1=\psi_4=0$, $\psi_0\neq0$,
	
	7) $\psi_0 = \psi_1=\psi_4=0$,
	
	8) $\psi_0 \neq 0, \ \psi_1\neq 0, \ \psi_4\neq 0$.
	\medbreak 
	The above class 7) is a trivial class consisting of null pinors.

	\section{Concluding remarks and outlook}
	
	In this work we implemented a systematic investigation of the normal, the almost complex, and the quaternionic  cases for the geometric Fierz identities, in the context of the Clifford bundles.   Using the geometric algebra approach of the Graf--Clifford, geometric  Fierz identities were employed to reconstruct the form-valued spinor and pinor bilinear covariants in certain dimensions and  signatures.
	New classes of spinors were constructed from the geometric Fierz identities for manifolds of signature $(1,2)$ and new classes of pinors have been derived, in spaces of signature $(9,0)$, besides constructive formul\ae\;
	to iteratively implement these constructions. 
	As in Ref. \cite{bab2} the K\"ahler--Atiyah algebra was constructed on metric cones and cylinders on $(9,0)$, the lifting of generalized Killing equations and the geometric Fierz
	isomorphisms can also be used to study 8-manifolds. 
	Beyond the scope of this paper, this last construction has a prominent link to the octonionic spinors and their classification in the context of the Lounesto's one \cite{Carrion:2003ve,Ablamowicz:2014rpa,Bengtsson:1987si}, and may be valuable for finding other solutions in string theory.  In fact, the new mass dimension one spinors found in Ref. \cite{Bonora:2017oyb} may be employed to construct pure spinor superstring  ghosts in a curved heterotic background \cite{Fleury:2015bsd}.

	\acknowledgments
	R.L.~is grateful to CAPES and R.dR.~is grateful to CNPq (Grant No. 303293/2015-2),
	and to FAPESP (Grant No. 2017/18897-8), for partial financial support.

	\appendix
	\section{Useful identities}
	Eqs. (\ref{eq123} -- \ref{eq124}) were disposed to construct new classes of pinors on signature $(9,0)$.
	Here we provide a detailed proof for each one of them. The notation and definitions ar the same thereon.
	\begin{enumerate}
		\item To derive Eq. (\ref{eq123}), since $\ \psi_0 \diamond \psi_0 = B(\alpha, \alpha) \psi_0$, then 
		\begin{eqnarray}\nonumber
			\psi_0 \ \blkdiam \  \psi_0 \!&\!=\!&\! 2 P_L (P_+ (\psi_0 \diamond \psi_0))= 2 P_L (P_+ (B(\alpha, \alpha) \psi_0))\nonumber\\&=& 2B(\alpha, \alpha) P_L ({\textstyle\frac{1}{2}} (\psi_0+ \star \psi_0))= 2B(\alpha, \alpha) P_L ({\textstyle\frac{1}{2}} (\psi_0+ \psi_0 \diamond \textbf{v}))\\ \nonumber
			&=& B(\alpha, \alpha)  ( \psi_0 + P_L(\underbrace{\psi_0 \wedge \textbf{v}}_{9-\text{form}}))\\
			&=&B(\alpha, \alpha)  \psi_0 \in \varGamma({{\textstyle \bigwedge}}^0).
		\end{eqnarray}
		\item Now, Eq. (\ref{eq1231}) follows from $ \ \psi_0 \diamond \psi_1 = B(\alpha, \alpha) \psi_1$, yielding 
		\begin{eqnarray}
			\ \psi_0 \ \blkdiam \  \psi_1 &=&  B(\alpha, \alpha) (\psi_1+ P_L(\psi_1 \diamond \textbf{v}))\nonumber\\&=& B(\alpha, \alpha)  ( \psi_1 + P_L( \underbrace{\psi_1 \wedge_1 \textbf{v}}_{8-\text{form}}))=B(\alpha, \alpha)  \psi_1 \in \varGamma({{\textstyle \bigwedge}}^1).
		\end{eqnarray}
		\item Analogously, the expression $\bullet \ \psi_0 \diamond \psi_4 = B(\alpha, \alpha) \psi_4$ implies that 
		\begin{eqnarray}
			\psi_0 \ \blkdiam \  \psi_4 &=&  B(\alpha, \alpha) (\psi_4+ P_L(\psi_4 \diamond \textbf{v}))
			= B(\alpha, \alpha)  ( \psi_4 + P_L(  {\textstyle\frac{1}{4!}}\underbrace{\psi_4 \wedge_4 \textbf{v}}_{5-\text{form}}))\nonumber\\&=&B(\alpha, \alpha)  \psi_4 \in \varGamma({{\textstyle \bigwedge}}^4),
		\end{eqnarray} showing, thus, Eq. (\ref{eq1232}).
		\item Next, Eq. (\ref{eq1233}) can be simply derived observing that $\psi_1 \diamond \psi_0 = B(\alpha, \alpha) \psi_1$, hence 
		\begin{eqnarray}
			\psi_1 \ \blkdiam \  \psi_0 =  B(\alpha, \alpha) (\psi_1+ P_L(\psi_1 \diamond \textbf{v})) = B(\alpha, \alpha)  \psi_1 \in \varGamma({{\textstyle \bigwedge}}^1).
		\end{eqnarray}
		\item Now, Eq. (\ref{eq1234}) follows from 
		$ \psi_1 \diamond \psi_1 = \sum^{1}_{r=0} \dfrac{(-1)^{r(1-r)+[\frac{r}{2}]}}{r!} \psi_1 \wedge_r \psi_1 = \psi_1 \wedge \psi_1 + \psi_1 \wedge_1 \psi_1 \in \varGamma ({{\textstyle \bigwedge}}^2) \oplus \varGamma ({{\textstyle \bigwedge}}^0)$, hence yielding 
		\begin{eqnarray}\nonumber
			\psi_1 \ \blkdiam \  \psi_1 &=& 2 P_L (P_+ ( \psi_1 \diamond \psi_1)) = 2 P_L (P_+ (\psi_1 \wedge \psi_1 + \psi_1 \wedge_1 \psi_1 ))\\ \nonumber 
			&=&  P_L (\psi_1 \wedge \psi_1 +\star(\psi_1 \wedge \psi_1) + \psi_1 \wedge_1 \psi_1+ \star (\psi_1 \wedge_1 \psi_1))\\ \nonumber
			&=& P_L (\psi_1 \wedge \psi_1 +(\psi_1 \wedge \psi_1)\diamond \textbf{v} + \psi_1 \wedge_1 \psi_1+ (\psi_1 \wedge_1 \psi_1)\diamond \textbf{v})\\ \nonumber
			&=& \psi_1 \wedge \psi_1 + \psi_1 \wedge_1 \psi_1  + P_L(-{\textstyle\frac{1}{2}}\underbrace{(\psi_1 \wedge \psi_1)\wedge_2 \textbf{v}}_{7-\text{form}} + \underbrace{(\psi_1 \wedge_1 \psi_1)\wedge \textbf{v}}_{9-\text{form}})\\
			&=&\psi_1 \wedge \psi_1 + \psi_1 \wedge_1 \psi_1 \in \varGamma({{\textstyle \bigwedge}}^2) \oplus \varGamma({{\textstyle \bigwedge}}^0).
		\end{eqnarray}
		\item Besides, Eq. (\ref{eq1235}) can be derived, starting from  the expression \begin{equation}\psi_1 \diamond \psi_4 = \sum^{1}_{r=0} \dfrac{(-1)^{r(1-r)+[\frac{r}{2}]}}{r!} \psi_1 \wedge_r \psi_4 = \psi_1 \wedge \psi_4 + \psi_1 \wedge_1 \psi_4 \in \varGamma ({{\textstyle \bigwedge}}^5) \oplus \varGamma ({{\textstyle \bigwedge}}^3).\end{equation} It then implies that 
		\begin{eqnarray}\nonumber
			\psi_1 \ \blkdiam \  \psi_4 &=& P_L (\psi_1 \wedge \psi_4 +(\psi_1 \wedge \psi_4)\diamond \textbf{v} + \psi_1 \wedge_1 \psi_4+ (\psi_1 \wedge_1 \psi_4)\diamond \textbf{v})\\ \nonumber
			&=&  \psi_1 \wedge_1 \psi_4  + P_L({\textstyle\frac{1}{5!}}\underbrace{(\psi_1 \wedge \psi_4)\wedge_5 \textbf{v}}_{4-\text{form}} -{\textstyle\frac{1}{3!}} \underbrace{(\psi_1 \wedge_1 \psi_4)\wedge_3 \textbf{v}}_{6-\text{form}})\\
			&=& \psi_1 \wedge_1 \psi_4 + \star(\psi_1 \wedge \psi_4) \in \varGamma({{\textstyle \bigwedge}}^3) \oplus \varGamma({{\textstyle \bigwedge}}^4).
		\end{eqnarray}
		\item By the expression  $\psi_4 \diamond \psi_0 = B(\alpha, \alpha) \psi_4$, Eq. (\ref{eq1236}) can be obtained:
		\begin{eqnarray}
			\psi_4 \ \blkdiam \  \psi_0 &=&  B(\alpha, \alpha) (\psi_4+ P_L(\psi_4 \diamond \textbf{v})) = B(\alpha, \alpha)  \psi_4 \in \varGamma({{\textstyle \bigwedge}}^4).
		\end{eqnarray}
		\item The penultimate case consists of starting from the equation \begin{equation}\psi_4 \diamond \psi_1 = (-1)^{1.4}\sum^{1}_{r=0} \dfrac{(-1)^{r(2-r)+[\frac{r}{2}]}}{r!} \psi_1 \wedge_r \psi_4 = \psi_1 \wedge \psi_4 - \psi_1 \wedge_1 \psi_4 \in \varGamma ({{\textstyle \bigwedge}}^5) \oplus \varGamma ({{\textstyle \bigwedge}}^3)\end{equation} to prove that 
		\begin{eqnarray}\nonumber
			\psi_4 \ \blkdiam \  \psi_1 &=& P_L (\psi_1 \wedge \psi_4 +(\psi_1 \wedge \psi_4)\diamond \textbf{v} - \psi_1 \wedge_1 \psi_4 - (\psi_1 \wedge_1 \psi_4)\diamond \textbf{v})\\ 
			&=& - \psi_1 \wedge_1 \psi_4 + \star(\psi_1 \wedge \psi_4) \in \varGamma({{\textstyle \bigwedge}}^3) \oplus \varGamma({{\textstyle \bigwedge}}^4),
		\end{eqnarray} which is Eq. (\ref{eq1236}).
		\item Finally, Eq. (\ref{eq124}) is now derived, from 
		\noindent $\psi_4 \diamond \psi_4 = \sum^{4}_{r=0} \dfrac{(-1)^{r(4-r)+[\frac{r}{2}]}}{r!} \psi_4 \wedge_r \psi_4 \\= \psi_4 \wedge \psi_4 - \psi_4 \wedge_1 \psi_4 - {\textstyle\frac{1}{2}} \psi_4 \wedge_2 \psi_4 + {\textstyle\frac{1}{3!}} \psi_4 \wedge_3 \psi_4 + {\textstyle\frac{1}{4!}} \psi_4 \wedge_4 \psi_4 \in \varGamma ({{\textstyle \bigwedge}}^8) \oplus \varGamma ({{\textstyle \bigwedge}}^6) \oplus \varGamma ({{\textstyle \bigwedge}}^4) \oplus \varGamma ({{\textstyle \bigwedge}}^2) \oplus \varGamma ({{\textstyle \bigwedge}}^0)$, implying that 
		\begin{eqnarray}\nonumber \ \psi_4 \ \blkdiam \  \psi_4 &=& P_L (\psi_4 \diamond \psi_4 + (\psi_4 \diamond \psi_4 ) \diamond \textbf{v}) \\ \nonumber
			&=& P_L (\psi_4 \wedge \psi_4\!-\!\psi_4 \wedge_1 \psi_4\!-\!{\textstyle\frac{1}{2}} \psi_4 \wedge_2 \psi_4\!+\!{\textstyle\frac{1}{3!}} \psi_4 \wedge_3 \psi_4\!+\!{\textstyle\frac{1}{4!}} \psi_4 \wedge_4 \psi_4\!+\!(\psi_4 \wedge \psi_4)\diamond \textbf{v}\\ \nonumber
			&& - (\psi_4 \wedge_1 \psi_4)\diamond \textbf{v} - {\textstyle\frac{1}{2}} (\psi_4 \wedge_2 \psi_4)\diamond \textbf{v} + {\textstyle\frac{1}{3!}} (\psi_4 \wedge_3 \psi_4)\diamond \textbf{v} + {\textstyle\frac{1}{4!}} (\psi_4 \wedge_4 \psi_4)\diamond \textbf{v})\\ \nonumber
			&=&  -\!{\textstyle\frac{1}{2}} \psi_4 \wedge_2 \psi_4\!+\!{\textstyle\frac{1}{3!}} \psi_4 \wedge_3 \psi_4\!+\!{\textstyle\frac{1}{4!}} \psi_4 \wedge_4 \psi_4 + P_L({\textstyle\frac{1}{8!}}\underbrace{(\psi_4 \wedge \psi_4)\wedge_8 \textbf{v}}_{1-\text{form}} \\ \nonumber
			&&+ {\textstyle\frac{1}{6!}}\underbrace{(\psi_4 \wedge_1 \psi_4)\wedge_6 \textbf{v}}_{3-\text{form}} - {\textstyle\frac{1}{4!}} {\textstyle\frac{1}{2}} \underbrace{(\psi_4 \wedge_2 \psi_4)\wedge_4 \textbf{v}}_{5-\text{form}} - {\textstyle\frac{1}{2}}{\textstyle\frac{1}{3!}} \underbrace{(\psi_4 \wedge_3 \psi_4)\wedge_2 \textbf{v}}_{7-\text{form}} \\ \nonumber
			&& + {\textstyle\frac{1}{4!}} \underbrace{(\psi_4 \wedge_4 \psi_4)\wedge \textbf{v})}_{9-\text{form}}\\ \nonumber
			&=&-\!{\textstyle\frac{1}{2}} \psi_4 \wedge_2 \psi_4\!+\!{\textstyle\frac{1}{3!}} \psi_4 \wedge_3 \psi_4\!+\!{\textstyle\frac{1}{4!}} \psi_4 \wedge_4 \psi_4\!+\!{\textstyle\frac{1}{8!}}(\psi_4 \wedge \psi_4)\wedge_8 \textbf{v}\!+\!{\textstyle\frac{1}{6!}}(\psi_4 \wedge_1 \psi_4)\wedge_6 \textbf{v} \\ \nonumber
			&=& -\!{\textstyle\frac{1}{2}} \psi_4 \wedge_2 \psi_4\!+\!{\textstyle\frac{1}{3!}} \psi_4 \wedge_3 \psi_4\!+\!{\textstyle\frac{1}{4!}} \psi_4 \wedge_4 \psi_4 + \star (\psi_4 \wedge \psi_4) - \star (\psi_4 \wedge_1 \psi_4)\\
			&& \in \varGamma({{\textstyle \bigwedge}}^4) \oplus \varGamma({{\textstyle \bigwedge}}^2) \oplus \varGamma({{\textstyle \bigwedge}}^0)\oplus \varGamma({{\textstyle \bigwedge}}^1)\oplus \varGamma({{\textstyle \bigwedge}}^3).
		\end{eqnarray}
	\end{enumerate}


\begin{thebibliography}{x}
		
		
		\bibitem{oxford}
		J.~Vaz Jr. and R. da Rocha,
		``An Introduction to Clifford Algebras and Spinors,'' Oxford Univ. Press, Oxford, 2016.
		
		\bibitem{Bonora:2009ta}
		L.~Bonora, F.~F.~Ruffino and R.~Savelli,
		\emph{Revisiting pinors, spinors and orientability,}
		Bollettino U. M. I. {\bf 9} IV (2012) [\href{https://arxiv.org/abs/0907.4334}{arXiv:0907.4334 [math-ph]}].
		
		\bibitem{lou2} P. Lounesto, \emph{Clifford Algebras and Spinors}, {Cambridge Univ. Press}, Cambridge, 2002.
		
		\bibitem{bilfierz} L. S. Randriamihamison, \emph{Identites de Fierz et formes bilineaires dans les espaces spinoriels}, J. Geom.
		Phys. {\bf{10}} (1992) 19.
		
		\bibitem{bab1}  C. I. Lazaroiu, E. M. Babalic and I. A. Coman,
		\emph{The geometric algebra of Fierz identities in arbitrary dimensions and signatures},   JHEP {\bf 1309} (2013) 156  [\href{https://arxiv.org/abs/1304.4403}{arXiv:1304.4403 [hep-th]}].
		
		\bibitem{Fabbri:2014zya}
		{\color{black}{  L.~Fabbri,
				\emph{Least-order torsion-gravity for dirac fields, and their non-linearity terms,} 
				Gen.\ Rel.\ Grav.\  {\bf 47} (2015)  1837
				[\href{https://arxiv.org/abs/1405.5129}{arXiv:1405.5129 [gr-qc]}].}}
		
		\bibitem{fabbri}	 L.~Fabbri,
		\emph{A generally-relativistic gauge classification of the Dirac fields}, 
		Int. J. Geom. Meth. Mod. Phys. {\bf 13} (2016) 1650078  [\href{https://arxiv.org/abs/1603.02554}{arXiv:1603.02554 [gr-qc]}].
		
		
		
		
		\bibitem{Fabbri:2014wda}
		{\color{black}{  L.~Fabbri, S.~Vignolo and S.~Carloni,
				\emph{Renormalizability of the Dirac equation in torsion gravity with nonminimal coupling}, 
				Phys.\ Rev.\ D {\bf 90} (2014)  024012
				[\href{https://arxiv.org/abs/1404.5784}{arXiv:1404.5784 [gr-qc]}].}}
		
		
		
		\bibitem{Vignolo:2011qt}
		S.~Vignolo, L.~Fabbri and R.~Cianci,
		\emph{Dirac spinors in Bianchi-I f(R)-cosmology with torsion},   J.\ Math.\ Phys.\  {\bf 52} (2011) 112502 
		[\href{https://arxiv.org/abs/1106.0414}{arXiv:1106.0414 [gr-qc]}].
		
		
		
		
		\bibitem{hoff}
		J.~M.~Hoff da Silva and R.~da Rocha,
		\emph{Unfolding Physics from the Algebraic Classification of Spinor Fields}, 
		Phys.\ Lett.\ B {\bf 718} (2013) 1519
		[\href{https://arxiv.org/abs/1212.2406}{arXiv:1212.2406 [hep-th]}].
		
		\bibitem{Rogerio:2017gvr}
		{\color{black}{  R.~J.~Bueno Rogerio, J.~M.~Hoff da Silva, M.~Dias and S.~H.~Pereira,
				\emph{Effective lagrangian for a mass dimension one fermionic field in curved spacetime,} 
				JHEP {\bf 1802} (2018) 145
				[\href{https://arxiv.org/abs/1709.08707}{arXiv:1709.08707 [hep-th]}]. }}
		
		\bibitem{Cra} J. P. Crawford, 
		\emph{On The Algebra Of Dirac Bispinor Densities:
			Factorization And Inversion Theorems},  
		J. Math. Phys. \textbf{26} (1985) 1439.
		
		\bibitem{Cavalcanti:2014wia}
		{\color{black}{ R.~T.~Cavalcanti,
				\emph{Classification of Singular Spinor Fields and Other Mass Dimension One Fermions,} 
				Int.\ J.\ Mod.\ Phys.\ D {\bf 23} (2014) no.14,  1444002 [\href{https://arxiv.org/abs/1408.0720}{arXiv:1408.0720 [gr-qc]}].}}
		
		\bibitem{Bon14}
		L.~Bonora, K.~P.~S.~de Brito and R.~da Rocha,
		\emph{Spinor Fields Classification in Arbitrary Dimensions and New Classes of Spinor Fields on 7-Manifolds,}
		JHEP {\bf 1502} (2015) 069 [\href{https://arxiv.org/abs/1411.1590}{arXiv:1411.1590 [hep-th]}].
		
		
		
		\bibitem{Bonora:2015ppa}
		L.~Bonora and R.~da Rocha,
		\emph{New Spinor Fields on Lorentzian 7-Manifolds}, 
		JHEP {\bf 1601} (2016) 133
		[\href{https://arxiv.org/abs/1508.01357}{arXiv:1508.01357 [hep-th]}].
		
		\bibitem{brito}
		K.~P.~S.~de Brito and R.~da Rocha,
		\emph{New fermions in the bulk,}
		J.\ Phys.\ A {\bf 49} (2016)  415403
		[\href{https://arxiv.org/abs/1609.06495}{arXiv:1609.06495 [hep-th]}].
		
		
		{\color{black}{
				
				
				\bibitem{Frohlich:1988qh}
				J.~Frohlich and P.~A.~Marchetti,
				\emph{Quantum Field Theories of Vortices and Anyons}, 
				Commun.\ Math.\ Phys.\  {\bf 121} (1989) 177.
				
				
				
				\bibitem{Grignani:1995vn}
				G.~Grignani, M.~Plyushchay and P.~Sodano,
				\emph{A Pseudoclassical model for P, T invariant planar fermions}, 
				Nucl.\ Phys.\ B {\bf 464} (1996) 189 [\href{https://arxiv.org/abs/hep-th/9511072}{arXiv:hep-th/9511072}].
				
				
				
				\bibitem{Gavrilov:2012jk}
				S.~P.~Gavrilov, D.~M.~Gitman and N.~Yokomizo,
				\emph{Dirac fermions in strong electric field and quantum transport in graphene,} 
				Phys.\ Rev.\ D {\bf 86} (2012) 125022
				[\href{https://arxiv.org/abs/1207.1749}{arXiv:1207.1749 [hep-th]}].
				
				
				
				
				\bibitem{Gonzalez:2009je}
				J.~Gonzalez and J.~Herrero,
				\emph{Graphene wormholes: A Condensed matter illustration of Dirac fermions in curved space,} 
				Nucl.\ Phys.\ B {\bf 825} (2010) 426
				[\href{https://arxiv.org/abs/0909.3057}{arXiv:0909.3057 [cond-mat.mes-hall]}].
				
				
				\bibitem{Dutreix:2013jva}
				C.~Dutreix, M.~Guigou, D.~Chevallier, C.~Bena,
				\emph{Majorana Fermions in Graphene and Graphene-Like Materials}, 
				Eur.\ Phys.\ J.\ B {\bf 87} (2014) 296
				[\href{https://arxiv.org/abs/1309.1143}{arXiv:1309.1143 [cond-mat.mes-hall]}].
				
				
				
				
				
				\bibitem{Mendes:2017hmv} 
				W.~M.~Mendes, G.~Alencar and R.~R.~Landim,
				\emph{Spinors Fields in Co-dimension One Braneworlds}, 
				JHEP {\bf 1802} (2018) 018 [\href{https://arxiv.org/abs/1712.02590}{arXiv:1712.02590 [hep-th]}].}}
		
		\bibitem{bab2} C. I. Lazaroiu, E. M. Babalic and I. A. Coman,
		\emph{Geometric algebra techniques in flux compactifications}, Adv.\ High Energy Phys.\  {\bf 2016} (2016) 7292534 [\href{https://arxiv.org/abs/1212.6766}{arXiv:1212.6766 [hep-th]}].
		
		
		\bibitem{frame}  R. Lopes, R. da Rocha,
		\emph{The Graf product: a Clifford structure framework on the exterior bundle}, to appear, Adv. Appl. Clifford Alg. {\bf 28} (2018) 57 [\href{https://arxiv.org/abs/1712.02737}{arXiv:1712.02737 [math.DG]}].
		
		
		\bibitem{cep}  T. Houri, D. Kubiz$\check{\rm n}$\'ak, C. Warnick, and Y. Yasui, \emph{Symmetries of the Dirac Operator with Skew-Symmetric Torsion}, 
		Class.\ Quant.\ Grav.\  {\bf 27} (2010) 185019.
		[\href{https://arxiv.org/abs/1002.3616}{arXiv:1002.3616 [hep-th]]}.
		
		
		\bibitem{Linhares:1985xa}
		C.~A.~Linhares and J.~A.~Mignaco,
		\emph{Su(4) For The Dirac Equation}, 
		Phys.\ Lett. B  {\bf 153} (1985) 82.
		
		\bibitem{Graf} W. Graf, \emph{Differential forms as spinors}, Annales de l'I. H. P. Physique th\'eorique {\bf 29} (1978) 85-109. [\href{https://eudml.org/doc/75997}{eudml:75997}].
		
		
		\bibitem{Car}  M.~Cariglia, P.~Krtous and D.~Kubiznak,
		\emph{Commuting symmetry operators of the Dirac equation, Killing-Yano and Schouten-Nijenhuis brackets},
		Phys.\ Rev.\ D {\bf 84} (2011)  024004. [\href{https://arxiv.org/abs/1102.4501}{arXiv:1102.4501 [hep-th]}].
		
		
		
		\bibitem{type} A. Van Proeyen, \emph{Tools for supersymmetry}, 2016 [\href{https://arxiv.org/abs/hep-th/9910030}{arXiv:hep-th/9910030}].
		\bibitem{oku1} S. Okubo, \emph{Real representations of finite Clifford algebras. I. Classification},  J. Math. Phys. {\bf 32} (1991) 1657.
		
		
		
		\bibitem{oku2} S. Okubo, \emph{Representations of Clifford Algebras and its applications}, Math. Jap. {\bf 41} (1995) 59 [\href{https://arxiv.org/abs/hep-th/9408165}{arXiv:hep-th/9408165}].
		
		\bibitem{Bil}  D.~V.~Alekseevsky, V.~Cortes, C.~Devchand and A.~Van Proeyen, \emph{Polyvector super-Poincar\'e algebras}, Commun.\ Math.\ Phys.\  {\bf 253}, 385 (2004). [\href{https://arxiv.org/abs/hep-th/0311107v2}{arXiv:hep-th/0311107}].
		
		\bibitem{Carrion:2003ve}
		H.~L.~Carrion, M.~Rojas and F.~Toppan,
		\emph{Quaternionic and octonionic spinors: A Classification}, 
		JHEP {\bf 0304} (2003) 040 [\href{https://arxiv.org/abs/hep-th/0302113}{arXiv:hep-th/0302113}]. 
		
		\bibitem{Ablamowicz:2014rpa}
		R.~Ablamowicz, I.~Gon\c calves and R.~da Rocha,
		\emph{Bilinear Covariants and Spinor Fields Duality in Quantum Clifford Algebras}, 
		J.\ Math.\ Phys.\  {\bf 55} (2014) 103501 [\href{https://arxiv.org/abs/1409.4550}{arXiv:1409.4550 [math-ph]}].
		
		
		
		
		
		
		
		
		\bibitem{Bengtsson:1987si}
		I.~Bengtsson and M.~Cederwall,
		\emph{Particles, Twistors and the Division Algebras,} 
		Nucl.\ Phys.\ B {\bf 302} (1988) 81.
		
		
		\bibitem{Bonora:2017oyb}
		L.~Bonora, J.~M.~Hoff da Silva and R.~da Rocha,
		\emph{Opening the Pandora's box of quantum spinor fields}, 
		Eur.\ Phys.\ J.\ C {\bf 78} (2018) 157 [\href{https://arxiv.org/abs/arXiv:1711.00544}{arXiv:1711.00544 [hep-th]}].
		\bibitem{Fleury:2015bsd}
		T.~Fleury,
		\emph{On the Pure Spinor Heterotic Superstring $b$ Ghost,} 
		JHEP {\bf 1603} (2016) 200 [\href{https://arxiv.org/abs/1512.00807}{arXiv:1512.00807 [hep-th]}].
		
	\end{thebibliography}
\end{document}